\documentclass[letterpaper,twocolumn,10pt]{IEEEtran}

\usepackage[utf8]{inputenc}
\usepackage{amssymb, amsmath}
\usepackage{xcolor,color,colortbl}
\usepackage{graphicx}
\usepackage{subcaption}
\usepackage{multirow}
\usepackage{comment}
\usepackage{array}
\usepackage{soul}
\usepackage{colortbl}
\usepackage{hhline}
\usepackage[normalem]{ulem}
\usepackage{amsthm}

\DeclareUnicodeCharacter{2212}{-}

\definecolor{myblue}{rgb}{0,0,1}

\begin{document}

\title{VERCEL: Verification and Rectification of Configuration Errors with Least Squares\\}

\author{Abhiram Singh, Sidharth Sharma and Ashwin Gumaste
\thanks{Abhiram Singh, Sidharth Sharma and Ashwin Gumaste are with the Department
of Computer Science and Engineering, Indian Institute of Technology Bombay, Mumbai-400076, India, 
(e-mail: abhiram25.1990@gmail.com, sidharth.sharma@ieee.org, ashwing@ieee.org).}}

\markboth{Accepted in IEEE/ACM Transactions on Networking}%
{Shell \MakeLowercase{\textit{et al.}}: Bare Demo of IEEEtran.cls for IEEE Journals}
%

\maketitle

\begin{abstract}
    We present Vercel, a network verification and automatic fault rectification tool that is based on a computationally tractable, algorithmically expressive, and mathematically aesthetic domain of linear algebra.
    Vercel works on abstracting out packet headers into standard basis vectors that are used to create a port-specific forwarding matrix $\mathcal{A}$, representing a set of packet headers/prefixes that a router forwards along a port.
    By equating this matrix $\mathcal{A}$ and a vector $b$ (that represents the set of all headers under consideration), we are able to apply \textit{least squares} (which produces a column rank agnostic solution) to compute which headers are reachable at the destination.  Reachability now simply means evaluating if vector $b$ is in the column space of $\mathcal{A}$, which can efficiently be computed using least squares.
    Further, the use of vector representation and least squares opens new possibilities for understanding network behavior. For example, we are able to map rules, routing policies, what-if scenarios to the fundamental linear algebraic form, $\mathcal{A}x=b$, as well as determine how to configure forwarding tables appropriately. We show Vercel is faster than the state-of-art such as  NetPlumber, Veriflow, APKeep, AP Verifier, when measured over diverse datasets. Vercel is almost as fast as Deltanet, when rules are verified in batches and provides better scalability, expressiveness and memory efficiency. A key highlight of Vercel is that while evaluating for reachability, the tool can incorporate intents, and transform these into auto-configurable table entries, implying a recommendation/correction system.
\end{abstract}


\section{Introduction}\label{sec:intro}
It is widely known that configuration errors in service provider networks form bulk of network outages \cite{zeng2012survey, bgp1, bgp2} resulting in operational challenges leading to conservative planning and slow rollout of services. 
Errors can be in the control or dataplane of routers, firewalls, middleboxes and switches. 
Control plane (protocol) verification \cite{cp1, Minesweeper, plankton, fayaz2016efficient, fast, cpCompress, VeriCon, assertion-lang} is harder to accomplish on account of the difficulties in abstracting the statefulness of protocols to computation models. 
In contrast, significant work exists in the realm of data plane verification \cite{dp1, static-reachability, libra, delta-net, veriflow, apkeep, ddnf, deltanet-icnp, mutable-datapaths, SymNet}, which is feasible (though complex).
The early work of atomic predicates \cite{atomic-predicate} involved mapping network forwarding rules with formal methods that resulted in a formulation whose solution led to answer reachablity verification queries.
The complexity of atomic predicates was further relaxed by the static HSA scheme \cite{hsa}, which considered packets in $L$-dimensional space (where $L$ is the number of bits in a packet header) and worked on tracking the transformation of such packets through multiple network boxes (routers, etc.) by applying and then conjoining transfer functions.
HSA was extended to include dynamic updates in NetPlumber \cite{netplumber}. 
Another work -- Veriflow \cite{veriflow}, mapped network-wide headers to equivalence classes (ECs) (defined as a set of packets that are treated similarly at routers) through a trie data structure. 
Veriflow was able to detect configuration errors in real-time.
Recent schemes APKeep \cite{apkeep} and Deltanet \cite{delta-net} improved the initial success of Veriflows' approach using optimizations and graph-theoretic means, resulting in faster reachability computation, loop-detection, blackhole identification (a packet at a node with no rule matching the packets' header). 

The next set of questions pertains to expressiveness, robustness, service support, and generic verification of network invariants. 
The solution to these lies in combining the initial successes of HSA/NetPlumber (the concept of $L$-dimensional space) with the EC approach by the more recent techniques (Veriflow, APKeep, Deltanet). 
To this end, our solution, \textit{Vercel}, applies linear algebraic techniques  -- a paradigmatic shift in the use of underlying techniques  towards solving the dataplane verification in real-time. Vercel uses both the dimensional transformations in HSA/Netplumber and ECs of Veriflow, and then applies linear algebra to achieve \textit{verification, recommendation and rectification}. 
To understand the rationale behind using linear algebra, recall the concept of header spaces in HSA/NetPlumber. 
Header spaces represent an $L$-bit header into an $L$-dimensional space and model a forwarding device as a function that transforms a header from one subspace to another. 
However, the complexity of transforming headers in subspaces is quadratic in the number of headers. 
In contrast, Vercel first identifies $m$ ECs by the rules collected from all the forwarding devices. 
Thereafter, Vercel creates an $m$-dimensional space corresponding to these ECs such that forwarding rules of each device form subspaces in this $m$-dimensional space.
Finally, to check reachability, Vercel models packet forwarding at routers by linearly projecting a point from one subspace to another along a path from the source to the destination. 
We argue that transforming a point (instead of subspaces as in HSA and NetPlumber) can best be done with linear algebra by converting ECs to vectors, and computing/tracing how these vectors move from a source to a destination, thus enabling us to evaluate for reachability. In addition to reachability, Vercel attains unprecedented scalability and efficiency (Section \ref{sec:evaluation}) as well as lays the groundwork for a more powerful abstraction -- that of providing a recommendation/rectification system.
It is established that linear projections can be efficiently handled by applying linear algebraic operations such as least squares \cite{strang}.
By definition, the linearity of our technique (linear algebra) implies speed of computation, availability of well-established theory, and readily available well-polished tools. 
Using these tools, we can verify network constructs and build a recommendation/rectification system that automatically rectifies configuration errors.
With Vercel's rectification technique, we can simply specify intents, and the tool automatically provisions tables, even when reachability does not seem to readily exist. It does so by computing the missing entries at tables and populating those, all in a single linear algebraic operation, which is possible only because we have transformed header spaces to vectors.
Vercel makes a jumpstart in auto-configuration, such that the tool recommends and corrects faults in both path selection and table manipulation.

To compute packet (or equivalently header) reachability between a source and destination router, Vercel initially creates a network-wide binary tree, as a data structure for efficiently representing headers present in the forwarding tables of routers.
This binary tree enables the division of the header space into non-overlapping partitions (similar to ECs), and thereafter, Vercel represents each partition as orthogonal vectors.
These unique vectors provide a foundation for introducing least squares.
With these vectors arranged in columns, Vercel creates a forwarding matrix  $\mathcal{A}$ for each port at a router.
Vercel initializes another vector $b$ for all the partitions whose reachability we desire to be evaluated. 
The matrix $\mathcal{A}$ and vector $b$ is then part of the standard linear equation $\mathcal{A}x=b$. 
The solution vector $x$ identifies which headers in $b$ are being forwarded along the port corresponding to $\mathcal{A}$.
We argue that if the matrix $\mathcal{A}$ contains orthogonal columns, then solving $\mathcal{A}x=b$ across different ports with least squares leads to solving for reachability. 
Applying least squares gives us excellent insights into computing reachability and fixing configuration faults. 

Another advantage of using linear algebra is that Vercel does away with checking reachability on multiple headers \textit{sequentially} (as in Veriflow and Deltanet); instead, it relies on vector spaces to apply a single algebraic operation simultaneously on multiple headers resulting in a key contribution of \textit{batch processing}, which is not supported by earlier schemes. Parallelism in other techniques can be obtained by running multiple threads, whereas Vercel takes advantage of CPU based (Table \ref{tab:whatif}) vector processing instructions even on a single thread, inducing scalability.
Vercel's key contributions are listed as follows:

1) \textit{Linear algebra to model networking devices}: 
    The merit of representing packet headers in a vector space is that it increases the expressive power beyond reachability. Vercel can be used to check loops, detect blackholes, confirm routing policies and answer \textit{what-if} scenarios in the presence of ACL rules and packet-header transformations.

2) \textit{Real-time network verification}: 
    We show that Vercel achieves significant improvement (on standard datasets, such as the one from Stanford \cite{stanford-datset}) of $8\times$ over Veriflow, $164\times$ over NetPlumber, $1.7\times$ over APKeep, $36\times$ over AP Verifier.
    Although Deltanet is faster than Vercel for verifying a single update, it cannot model a diverse set of network functions such as NAT and ACL and consumes up to $7.8\times$ more memory as compared to Vercel. Also, in the case of batch updates, the performance of Vercel is comparable to Deltanet (which does not implicitly even support batch processing).
    
3) \textit{Scalability}: Vercel performs well on large networks such as a synthetic network of 2000 nodes with millions of rules. 
Even with a network of this size, the verification time is of the order of 500$\mu$s.

4) \textit{Using linear algebra for recommendation and rectification}: Linear regression being a by-product of least squares, can also be used as a recommendation system (how well are forwarding tables configured). 
    The standard equation $\mathcal{A}x=b$ may not have ``\textit{a}" solution (because $\mathcal{A}$ may not be a full rank matrix) and the error resulting from least squares (Chapter 4 of \cite{strang}) tells us the efficiency of routing, as well as also paves the way for automatically provisioning intents and rectifying errors.

Network verification is important to avoid expensive outages. For example, in 2021, the Meta (Facebook’s) network went down globally due to a BGP configuration fault \cite{facebook-outage}. 
Configuration faults are the cause of 70 percent of global network outages, and this motivates approaches towards larger coverage of configuration fault detection. Our approach Vercel is different in treatment compared to other solutions – while other solutions use techniques such as atomic predicates or transformations, or SAT solvers, our approach is based on least-squares, exploring a paradigm never done before. 
This approach offers multiple advantages. It helps us verify configurations, detect misconfigurations, and develop a recommendation system that automatically rectifies errors.
This fundamental step forward (recommendation and rectification), in our view, sets Vercel apart from the rest of the verification solutions.  

\textit{Paper organization}: Section \ref{sec:overview} provides an overview of Vercel, while functional blocks of Vercel are presented in Section \ref{sec:design}. Section \ref{sec:reduce_time} provides an optimized implementation of Vercel. Section \ref{sec:beyond} extends Vercel to a variety of network functions while evaluation of Vercel is presented in Section \ref{sec:evaluation}. We discuss related works in Section \ref{sec:related}. Finally, Section \ref{sec:conclude} concludes the paper.

\section{Vercel Overview}\label{sec:overview}

\begin{figure*}[t]
    \centering
    \begin{subfigure}[t]{0.5\linewidth}
        \includegraphics[width=0.95\textwidth]{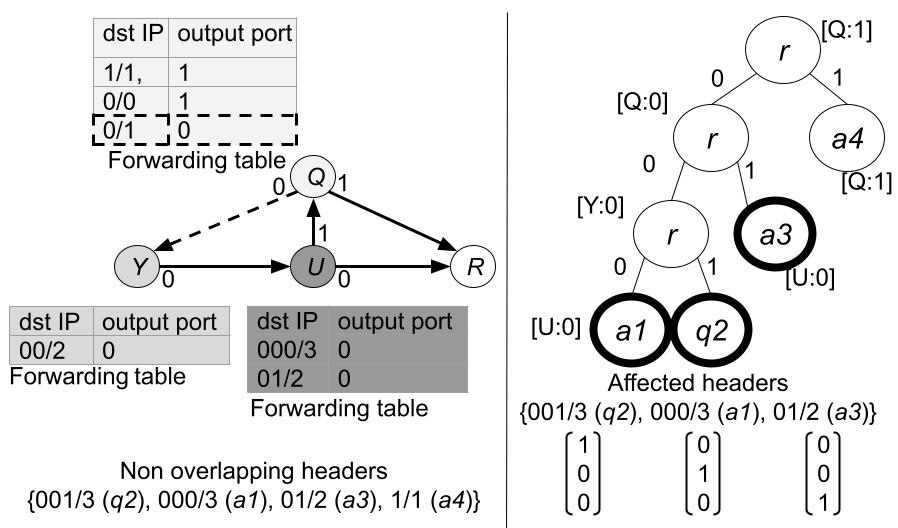}
        \centering
        \caption{}
        \label{fig:toynet_new}
    \end{subfigure}%
    \begin{subfigure}[t]{0.5\linewidth}
        \includegraphics[width=0.95\textwidth]{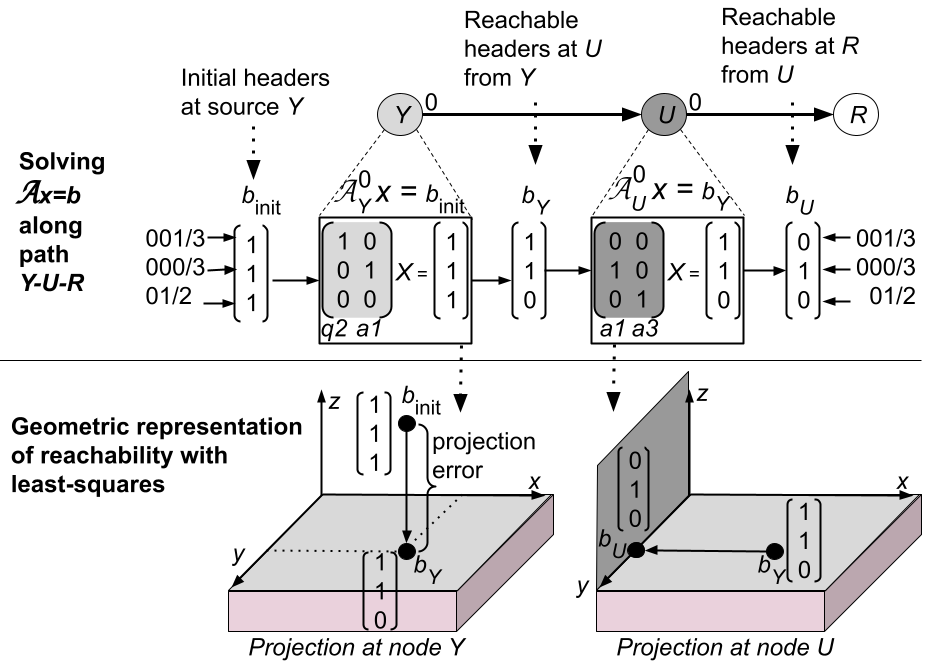}
        \centering
        \caption{}
        \label{fig:projection_matrices}
    \end{subfigure}
    \caption{(a) A toy network of 4 routers in which a new rule at router $Q$ is being inserted in its forwarding table (shown in dashed block). Vercel represents all rules in a binary tree and after insertion of a new rule, identifies 3 headers whose reachability might be affected. (b) An example to demonstrate reachability between router $Y$ and $R$ along the path $Y-U-R$ with least squares. Vercel represents 3 headers in 3-dimension with the orthogonal vectors. Forwarding rules of routers $Y$ and $U$ are represented in $x$-$y$ and $y$-$z$ planes, respectively. Vercel computes reachability by sequentially projecting a point in 3-dimension to the subspace created for router $Y$ and $U$.}
    \label{fig:projection_example}
\end{figure*}

\begin{table*}[t]
\caption{Symbols and notations used.}
\label{tab:notations}
\begin{center}
\scriptsize
\begin{tabular}{ 
|c|p{4cm}|c|p{4cm}|c|p{4cm}|}
\hline
\rowcolor{lightgray}
Notation & Definition & Notation & Definition & Notation & Definition\\ 
\hline
$L$ & Size of header in bits & $m$	& Number of equivalence classes & $N((\mathcal{A}_i^p)^T))$ & Null space of $\mathcal{A}_i^p$  \\
\hline
$r$	& Supernet headers& $a$ 	& Atomic headers & $C(\mathcal{A}_i^p)$ & Column space of $\mathcal{A}_i^p$   \\
\hline
$q$ & Represents iatomic headers  & $[\mathcal{A}|b]$ &  Augmented matrix & $H$ & Represents the unit step function \\
\hline
$\mathcal{A}$ &  A set of packet
headers/prefixes that a router forwards along a port  & $b$  &  Represents the set of all headers under consideration & $g_i$ & An $m$-dimensional filtering vector with binary values (0/1) \\
\hline
$x$	& The solution vector for linear equations $\mathcal{A}x=b$  & $n$	& Headers forwarded along a port, $1 \leq n\leq m$ & $\hat{x}$	& Approximate solution using least squares \\ 
\hline
$\mathcal{A}_Y^0$ & Headers forwarded by router $Y$ on its port $0$ & $S\textsuperscript{affected}$ &  Set of  headers whose reachability may have been affected  &  $T_i$ & Transformation matrix  \\ 
\hline
$f^{acl}$ & Filtering function that maps a set of packet headers to a set of actions  &  $v_i^p$ & Atomic+iatomic headers (in $S\textsuperscript{affected}$) that router $i$ forwards to its port $p$ & $f^{tr}$ & A set of rules containing the header ``match" and ``transformed" fields \\
\hline
$P\textsuperscript{affected}$ & A set contains those ports on which a router forwards the packet with the header in the set $S\textsuperscript{affected}$ & $c_i$ & The indices with a non-zero entry in it represent packets received at router $i$ that do not get any match in the forwarding table & 
$t_i$ & Represents atomic+iatomic headers reachable up to the router $i$ and then transformed using $T_i$
\\
\hline
\end{tabular}
\vspace{-5pt}	
\end{center}
\end{table*}

This section presents a high-level overview of Vercel, including an intuition for applying linear algebra for network verification. Notations used in the paper are listed in Table \ref{tab:notations}.
Vercel actively 'listens' to  configuration updates (either made through an SDN controller or any  other tool) to capture topology  and forwarding tables/updates.
Working on the granularity of individual packet headers is infeasible; hence Vercel, on the same lines of  \cite{plankton, delta-net, veriflow, apkeep}, groups headers so that packets in the same group are similarly dealt with across 
routers.

To group packet headers, we need to parse rules from all the routers.
Solely for the purpose of storing rules, we create a binary tree, which preserves the hierarchical representation of packet headers (Section \ref{subsec:bintree}, \ref{subsec:nodesbintree}).
For every header, the tree enables a path starting from the root towards the leaves denoting header bits.
The use of binary tree is similar to the use of \textit{trie} by Veriflow.
However, unlike Veriflow, where the leaves contain node-rule pairs, in our case, leaves imply non-overlapping headers (ECs). 
Since Vercel is primarily designed to verify invariants in real-time; therefore, Vercel traverses the binary tree to determine $m$ headers, whose reachablity might have been altered after a rule update (Section \ref{subsec:affectedbintree}).
Thereafter, Vercel uniquely maps $m$ headers to $m$-orthogonal vectors that collectively create an $m$-dimensional space.
After creating the $m$-dimensional space, Vercel identifies a subspace for each port on which a router forwards $1 \leq n\leq m$ headers.
After obtaining $n$ headers, Vercel defines a subspace for each port by selecting $n$ corresponding orthogonal vectors and storing these vectors in a matrix ($\mathcal{A}$) of dimension $(m, n)$ (Section \ref{subsec:subspaces}).
In addition to matrix $\mathcal{A}$, Vercel initializes an $m$-dimensional vector $b$, which is designed to evaluate the reachability of all $m$ headers under consideration. 

Matrix $\mathcal{A}$ and vector $b$ come together to form standard linear equation $\mathcal{A}x=b$, whose solution will eventually confirm if reachability exists (Section \ref{subsec:leastsq}).
Specifically, we solve $\mathcal{A}x=b$ for each port along the path between a source and destination.
Three cases exist while solving $\mathcal{A}x=b$.

In the first case, there exist rules that forward some but not all of the received packets to the specified port.
This implies vector $b$ is \ul{not} in the column space of matrix $\mathcal{A}$, indicating that no solution can be found for $\mathcal{A}x=b$.
Therefore, Vercel finds an approximate solution $\hat{x}$ using least squares.
For this, Vercel obtains a projection point $\mathcal{A} \hat{x}$ in the column space of the matrix $\mathcal{A}$.
The approximate solution vector $\hat{x}$ selects those headers in $b$ that the router forwards along its output port.

In the second case, if there exist forwarding rules for all the packets received at the router, then vector $b$ \ul{is in} the subspace defined by the columns of matrix $\mathcal{A}$.
The solution vector $x$ in $\mathcal{A}x=b$ points to \textit{all} the headers in $b$ that are forwarded along the port. 

In the third case, if a router forwards none of the received packets along the selected port, then vector $b$ lies \ul{in the null space} of matrix $\mathcal{A}^T$ and least squares returns a solution $\hat{x}=0$.

It is possible to efficiently solve all three cases. 
We argue that if matrix $\mathcal{A}$ contains orthogonal columns, then solving $\mathcal{A}x=b$ across different ports with least squares provides the projection of $b$ in the intersection of subspaces and finds a solution to the reachability problem (Section \ref{subsec:leastsq}). 

If $\mathcal{A}$ was a full rank matrix, the solution would be easy, but since that is not always the case, 
the next best thing to do is apply least squares to obtain a projection point.

Note that solving $\mathcal{A}x=b$ with options like matrix inversion, row reduction or linear optimization, is not efficient:
(1) Finding the inverse of matrix $\mathcal{A}$ to solve $\mathcal{A}x=b$ is not feasible as $\mathcal{A}$ is not of full rank ($m\neq n$).
Computing pseudo-inverse to obtain $x=\mathcal{A}^+b$ needs singular value decomposition (SVD), (a multi-step process), infeasible in real-time. (2) Row reduction algorithms are slow as they are cubic in time. (3) Linear optimization results in  infeasibility (because vector $b$ may not be present in the column space of $\mathcal{A}$.

In contrast, least squares guarantees to provide a solution of $\mathcal{A}x=b$ irrespective of the size of $\mathcal{A}$ and nature of $b$. 
Note that with the orthogonality condition imposed on the columns of matrix $\mathcal{A}$, we can find the projection of $b$ in the column space of $\mathcal{A}$ in linear time.
Importantly, note that though we use least squares to compute reachability, the perceived approximation characteristic of least squares has no bearing on the exactness/correctness of the reachability calculation. 
In the classical least squares model, error plays a role, and it may give the reader the impression that such error may lead to approximation, which is definitely not the case with Vercel.
For example, consider the first case, where least squares finds an approximate solution $\hat{x}$.
Though $\hat{x}$ is said to be an approximate solution, however it only selects a subset of headers in $\mathcal{A}$, specifically those for which forwarding rules are present at the said router.
In any case, $\hat{x}$ will not select false positives/negatives because of the approximation.

Moving forward, intuitively, least squares projects $b$ at the intersection of subspaces corresponding to the ports present along a path from the source to the destination.
The intersection of subspaces corresponds to the intersection of headers generated from the rules at routers.
The representation of headers in vector spaces enables Vercel to apply linear algebraic operations to simultaneously process headers in a single step, thereby providing speed up in verification.

\subsection{Example}
We now illustrate the functionality of Vercel with an example. Figure \ref{fig:toynet_new} is a toy network of 4 routers.
For simplicity, we consider 3-bit headers though generalization to more bits is trivial.
In this example, Vercel checks for reachability between source router $Y$ and destination $R$. 
The forwarding table at router $U$ shows that packets with header 000/3 and 01/2 are forwarded along output ports $0$ and $1$, respectively.
Vercel collects rules from the forwarding tables and utilizes a binary tree to arrange packet headers (as in Figure \ref{fig:toynet_new}).
While doing so, Vercel also keeps track of (router, port) pairs for each rule. 
The 4 leaf nodes of the binary tree $\{a1,\;q2,\;a3,\;a4\}$ denote non-overlapping headers $\{000/3,\; 001/3,\; 01/2,\; 1/1\}$.
In the figure, the node labels, i.e., $r$ (supernet), $a$ (atomic) and $q$ (iatomic) denote a specific class of headers defined later in Section \ref{subsec:nodesbintree}.

Now assume a network administrator inserts a new rule at router $Q$ (as shown in Figure \ref{fig:toynet_new}), which forwards packets with header 0/1 on its output port $0$ (shown by dashed block in the forwarding table at router $Q$).
Vercel detects the rule update and inserts the corresponding header 0/1 in the binary tree.
Post insertion, Vercel now traverses the headers in its subtrees to check if any of them were impacted for reachability during the update. 
There are three such (impacted) headers i.e., 000/3, 001/3, 01/2, which are added to a set, $S\textsuperscript{affected}$.
Vercel creates 3-dimensional orthogonal vectors [1, 0, 0], [0, 1, 0] and [0, 0, 1]  corresponding to impacted headers 001/3, 000/3 and 01/2.
For simplicity, we have selected orthogonal vectors with binary values.
Next, Vercel creates a vector $b_{init}=$ [1, 1, 1] that corresponds to three non-overlapping headers $\{q2,\;a1,\;a3\}$ for which we want to evaluate reachability.
After creating $b_{init}$, Vercel creates subspaces for port 0 at routers $Y$, $U$ and $Q$.
This is because along port 0, these routers forward some packets whose headers are in the set $S\textsuperscript{affected}$.
Figure \ref{fig:projection_matrices} shows $x$-$y$ and $y$-$z$ planes corresponding to the $(router, port)$ pairs $\{(Y, 0),\; (U, 0)\}$, forming two subspaces.
Since router $Y$ forwards packets with header 001/3 and 000/3 on its port $0$, therefore its subspace ($x$-$y$ plane) is defined by a matrix $\mathcal{A}_Y^0$.
Similarly, router $U$ forwards packets with header 000/3 and 01/2 on its port $0$, therefore its subspace ($y$-$z$ plane) is defined by a matrix $\mathcal{A}_U^0$.

After creating the subspaces for different ports, Vercel evaluates reachability by modelling packet forwarding at router $Y$.
This is done by solving $\mathcal{A}_Y^0x=b_{init}$.
At router $Y$, vector $b_{init}$ neither lies in the column space ($C(\mathcal{A}_Y^0)$) nor in the null space of $(\mathcal{A}_Y^0)^T$ ($Y$ in superscript denotes the matrix transpose).
Equivalently, $b_{init}$ lies at the union of the column and null space of $\mathcal{A}_Y^0$ and $(\mathcal{A}_Y^0)^T$, respectively. 
The above scenario implies that no solution is possible for $\mathcal{A}_Y^0x=b_{init}$. 
In this case, least squares finds an approximate solution resulting in a projection point $b_Y=$ [1, 1, 0] in $C(\mathcal{A}_Y^0)$, i.e., $x$-$y$ plane.
The point $b_Y$ implies that router $Y$ forwards packets with headers 001/3 and 000/3 through port 0, which are then received by router $U$ on path $Y-U$.
Since packets with header 01/2 are not forwarded from router $Y$, therefore the third index (corresponding to header 01/2) of vector $b_Y$ contains a 0 entry.
Similarly, Vercel models the packet forwarding at router $U$ at port $0$ by solving $\mathcal{A}_U^0x=b_Y$ and resulting in a projection point $b_U=$ [0, 1, 0] in the $y$-$z$ plane, indicating that router $U$ only forwards packets with header 000/3 through port $0$.
Since router $R$ receives a non-zero vector $b_U$ from router $U$, therefore Vercel confirms reachability between router $Y$ and $R$ along the path $Y-U-R$.
Finally, at the destination (router $R$), Vercel determines the set of reachable packet headers by computing the dot product between $b_U$ and vector representation of headers in the set $S\textsuperscript{affected}$.
For this example, Vercel obtains a non-zero dot product between $b_U$ and vector [0, 1, 0] (corresponding to the header 000/3), implying packets with header 000/3 are reachable from router $Y$ to $R$.
\begin{figure}[t]
    \centering
    \includegraphics[width=\linewidth]{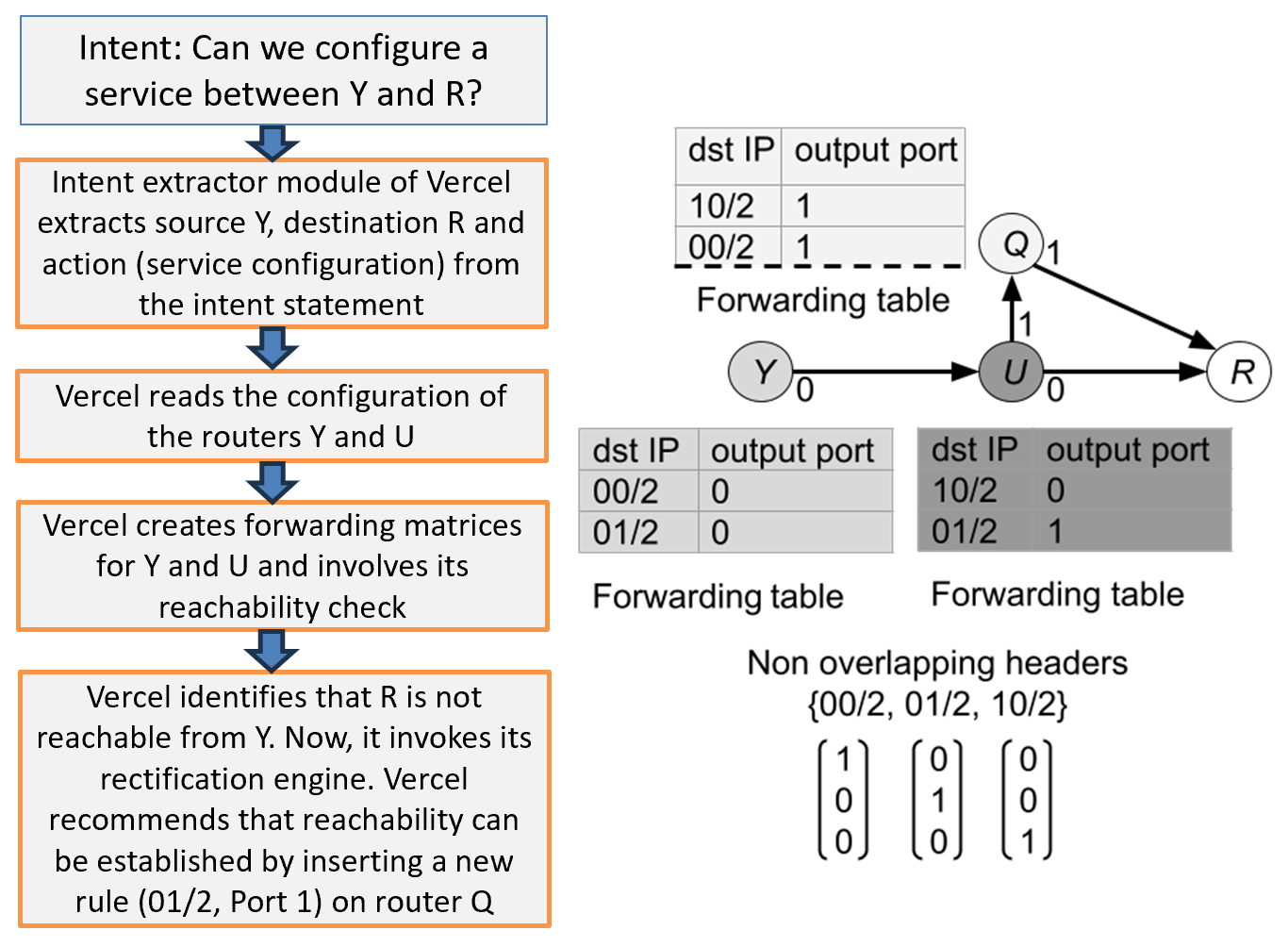}
    \caption{Steps for Vercel’s handling of user-specified intent related to service configuration (in the network shown on the right side of the figure).}
    \label{fig:intent}
    \vspace*{-10pt}
\end{figure}

\subsection{Rectification and Automatic Provisioning}\label{subsec:rect_prov_recomm}

Using least squares, we check for reachability and express intent to establish a service between two nodes in the network.
If reachability between these two nodes does not exist, then Vercel recommends a path (Section \ref{subsec:pathreco}) and rectifies the situation by \textit{automatically} populating tables at interim routers leading to reachability. 
The fact that misconfigurations are not just found, but also rectified and that the rectification is a direct by-product of $\mathcal{A}x=b$ using least squares, is a significant value addition to operators.
An operator only has to express a high level intent for a bunch of services and not worry about how these are correctly provisioned. The correctness is taken care of by Vercel.

Let us assume we performed least squares on $\mathcal{A}x=b$ for a port and identified that our target vector $b$ is in the null space of $\mathcal{A}^T$. Which means ``all" the headers represented by $b$ are not forwarded along that port.  Now, if we were to ensure reachability for any of the headers in $b$, then $b$ must be in the column space of $\mathcal{A}$. 
The simplest way of doing this is to replace $\mathcal{A}$ with an augmented matrix $[\mathcal{A}|b]$.
\begin{figure}[t]
    \centering
    \includegraphics[width=\linewidth]{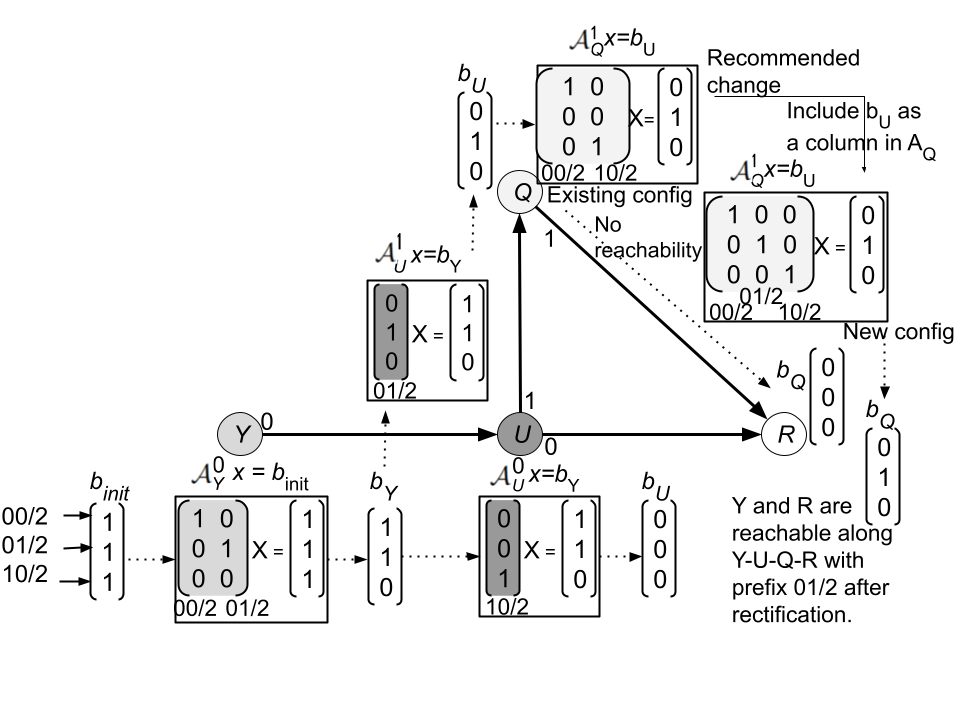}
    \caption{Example demonstrating internals of Vercel's rectification system. Initially, $R$ is not reachable from $Y$. 
    However, after applying Vercel's rectification at $Q$, reachability till $R$ is ensured along the path $Y-U-Q-R$.
}
    \label{fig:rectification}
    \vspace*{-10pt}
\end{figure}

We now describe rectification via the following example resulting from a users' intent. 
Consider a four-node network consisting of routers $Y$, $U$, $Q$ and $R$, shown on the right side of Fig. \ref{fig:intent}. 
Forwarding tables at each of the routers and corresponding vector notations for their non-overlapping headers are also shown in the figure. 
Assume an intent by the user whose goal is to set up a service between nodes $Y$ and $R$.
Vercel is responsible for parsing the intent statement to extract information such as source ($Y$), destination ($R$), and action (service provisioning).
Vercel then examines the configuration of routers $Y$ and $U$, generates forwarding matrices for both, and initializes $b_{init}$ (all prefixes) for reachability check. 
Upon analysis, Vercel determines that destination $R$ is currently unreachable from source $Y$ (as there are no forwarding rules installed at routers $U$ and $Q$ to forward packets received from $Y$ to $R$). 
In response, Vercel enables its rectification engine and proposes a solution based on linear algebraic operations. 
The activated rectification engine creates a new rule (01/2, Port 1) at router $Q$ to establish reachability.
The verification and rectification engines of Vercel work together for this intent, and the flow of operations is shown in Fig. \ref{fig:intent}. A detailed explanation of the linear algebraic operations performed for rectification is provided below.

In Fig. \ref{fig:rectification} is an example demonstrating Vercel's rectification system for the intent expressed in Fig. \ref{fig:intent}. 
We apply Vercel to verify reachability, for which matrices $\mathcal{A}$ and vector $b$ are created as shown in Fig. 
\ref{fig:rectification}. 
After applying $\mathcal{A}x=b$ at router $Y$ and $U$, we get $b_U$, which is a set of prefixes reachable from $Y$ to $Q$ (through path $Y-U-Q$). 
Now at router $Q$, while applying $\mathcal{A}_Qx=b_U$, Vercel finds that $b_U$ is in the null space of $\mathcal{A}_Q^T$, which implies none of the prefixes in $b_U$ will be forwarded towards $R$. 
At this point, our recommendation system kicks in and suggests a ``fix" to address the fault at router $Q$, such that reachability between $Y$ and $R$ is established.
As part of the fix, Vercel’s recommendation system adds $b_U$ as a new column of matrix $\mathcal{A}_Q$ (and updates the corresponding forwarding rules at router $Q$). 
Now, the updated matrix $\mathcal{A}_Q$ has three columns. This solves the problem as when we evaluate $\mathcal{A}_Qx=b_U$, $b_U$ will be in the column space of $\mathcal{A}_Q$. As a result of $\mathcal{A}_Qx=b_U$, we get $b_Q= [0,1,0]$, which is a vector that represents the prefixes reachable till destination $R$. Since this vector is now non-zero, a few prefixes are now reachable at the destination after applying the fix.

Nevertheless, employing rectification to address issues may inadvertently result in unintended consequences, such as rendering some destinations unreachable. To mitigate this, our rectification system operates based on the following principle: when implementing a fix (by adding a rule), the reachability of any current header should remain unaffected. For this, identifying the appropriate header is crucial whenever a rule update is conducted as part of rectification. In this regard, Vercel opts for a header that was previously unreachable between the source and destination pairs of interest. 
In accordance with this principle, the following procedure is followed during the rectification process.

Assume the rectification system suggests adding a rule with header $z$ in one of the routers’ forwarding tables.
Now, Vercel calculates if there exist any reachable headers $z’$ (for some other source-destination pairs) in the same table that overlap with $z$. 
Vercel then performs $z-z’$ to check if the resultant set is non-empty.
The non-emptiness here implies the existence of the non-reachable headers between source and destination. 
If so, it configures the corresponding non-empty set as a separate rule in the forwarding table. 
In case the result ($z-z’$) is empty, then Vercel attempts to find a header (other than $z$) and repeats the process. If we exhaust all possible `z’s through computation and still cannot find a non-empty $z-z’$, then Vercel notifies that rectification is not possible. By computing such $z-z’$, Vercel ensures that while adding a rule during the rectification process, it does not impact the reachability of any existing header.

\section{Vercel: Under the hood}\label{sec:design} 
This section describes the detailed working of Vercel. 

\subsection{Representing headers in a binary tree}\label{subsec:bintree}

\begin{figure}[t]
    \centering
    \includegraphics[width=\linewidth]{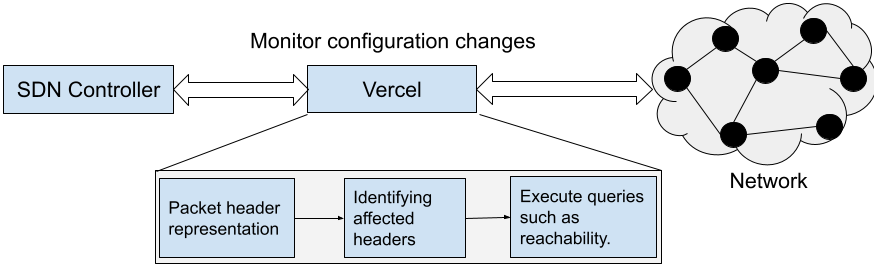}
    \caption{Vercel components.}
    \label{fig:netver_components}
\end{figure}

Initially, Vercel queries routers or an SDN controller to obtain forwarding tables and compute the network topology as shown in Fig. \ref{fig:netver_components}.
Thereafter, Vercel arranges all the existing packet headers (extracted from rules in the forwarding tables) in the form of a binary tree.
For adding a new header, Vercel first identifies its correct position in the tree by traversing the tree based on $L$-header bits.
These bits form a path from the root to a prospective node, where the new header is to be inserted.  
At every level (in the tree) the corresponding header bit indicates the traversal direction (0 = left and 1 = right branch). 
A path from the root to an intermediate/leaf node corresponds to a specific packet header.
The role of the binary tree  is in efficiently dividing the packet header space into non-overlapping partitions.
Each packet belonging to a partition observes similar processing at all the routers in the network.
The binary-tree enabled succinct representation provides Vercel a way in dividing the packet header space into three categories: \textit{supernet}, \textit{atomic} and \textit{iatomic}. 
Out of these, \textit{atomic} and \textit{iatomic} categories together represent all the non-overlapping headers extracted from the forwarding tables across the network, while \textit{supernet} represents the union of one or more \textit{atomic} and \textit{iatomic} headers.

\subsection{Classifying the nodes of the binary tree}\label{subsec:nodesbintree}
Vercel labels a packet header as a \textit{supernet}, if the corresponding path snippet from the root terminates at a non-leaf node.
The remaining packet headers extracted from the rules (whose corresponding paths from the root terminate at a leaf node) are denoted as \textit{atomic}.
Due to the hierarchy in the header representation, multiple headers (e.g., supernet and atomic) may overlap. 
Therefore, Vercel further fragments supernets to create a new category of headers (called \textit{induced-atomic headers} (\textit{iatomic}) that do not overlap with the atomic headers.
For example, in Figure \ref{fig:toynet_new} we observe 3 atomic headers (labeled as $a$), 1 iatomic header (labeled as $q$) and 3 supernets (labeled as $r$) in the binary tree.
Note that the atomic and iatomic headers represent the leaf nodes of a binary tree.

In addition to the attribute of label types ($r/a/q$), each node of the binary tree is also characterized by a list containing tuples.   
A tuple $(i,\;p)$ at node $k$ denotes that router $i$ contains a rule that maps a packet header (represented by a path from the root to node $k$) to an output port $p$.
In addition, Vercel assigns an integer identifier to each leaf node that distinguishes headers within the set of atomic and iatomic headers.

So far, we have defined atomic and iatomic headers; however, a crucial aspect is to analyze the task of identifying and segregating them once the tree is constructed. 
The following Lemma and Theorem provide a theoretical understanding of the procedure and complexities involved in determining the atomic and iatomic headers (the reader may skip the proofs for continuity). 


\textbf{Lemma 1}: The worst-case space and time complexity of finding all atomic headers in a network containing $m$ forwarding rules is $\mathcal{O}(m)$. 

\textit{Proof of Lemma 1:}
In the worst case, $m$ forwarding rules are non-overlapping with each other, and therefore, corresponding headers represent leaf nodes of the binary tree.
In this case, Vercel labels all leaf nodes ($m$) as atomic by performing a post-order traversal of the tree, implying that the space complexity of the atomic headers is $\mathcal{O}(m)$.
A post-order traversal of the binary tree with $m$ leaf nodes can be performed in $\mathcal{O}(m)$ time. Therefore, the time complexity of identifying atomic headers is also $\mathcal{O}(m)$.

\textbf{Theorem 1}: Given $m$ forwarding rules and a set of atomic headers having space complexity of $\mathcal{O}(m)$, the space complexity of the smallest size induced-atomic set is $\mathcal{O}(m)$ and it can be computed with the time complexity of $\mathcal{O}(m)$.


\textit{Proof of Theorem 1:}
\textit{Space complexity of iatomic nodes}: Consider a general case, in which multiple (say $u$) leaf nodes (atomic headers) have an ancestor supernet ($r$) with a maximum path length of $v$ between a supernet and an atomic header.
To represent all headers of the supernet ($r$) using non-overlapping partitions, Vercel introduces a maximum of $uv$ number of new iatomic headers in the tree along the path between $r$ and $v$. 
In the extreme case, when a supernet is a root node of the binary tree and all other headers are atomic, the number of iatomic nodes cannot be greater than $\mathcal{O}(mw)$, where $w$ is the maximum path length between the root node and a leaf node and $m$ is the number of rules in the forwarding tables across the network. 
A similar argument holds for multiple supernets having a common ancestor (as a supernet) in the binary tree.
Since $w$ is dependent on the header length (e.g., an IP prefix has a maximum value of 32 in case of IPv4) and $w\ll m$, therefore $w$ is treated as a constant and space complexity of iatomic headers becomes $\mathcal{O}(m)$.

\textit{Time complexity of adding iatomic nodes:}
Vercel creates all iatomic headers (new leaf nodes) by performing the post-order traversal of the binary tree.
While doing so, Vercel checks whether the current node of the tree is labeled as a supernet.
If the current node is labeled as a supernet, then for all descendant nodes having a single child, Vercel creates a second child and labels it as iatomic (constant time operation). 
Since the binary tree has the space complexity $\mathcal{O}(m)$ (with $m$ rules and $\mathcal{O}(m)$ atomic headers); therefore, the time complexity of post-order traversal along with the addition of new leaf nodes (iatomic headers) is $\mathcal{O}(m)$.

\subsection{Identifying affected atomic+iatomic headers post a rule update}\label{subsec:affectedbintree}
Vercel is always tuned into the network to intercept any rule update at the routers.
Once Vercel identifies an update, it computes whether reachability between different source-destination pairs is intact.
For this, Vercel  determines corresponding atomic+iatomic headers from the created binary tree, whose reachability may be affected.
Vercel denotes all such headers in a set $S\textsuperscript{affected}$.
To create $S\textsuperscript{affected}$, Vercel traverses a path of length $L$ from the root to a node $o_j$, directed by the $L$ bits of the header (specified by the rule being inserted/deleted). 
Thereafter, Vercel traverses subtrees of node $o_j$.
During this traversal, if Vercel observes an atomic/iatomic node, then it appends the node's identifier in the set $S\textsuperscript{affected}$. Besides the set $S\textsuperscript{affected}$, Vercel creates another set $P\textsuperscript{affected}$, which contains those ports on which a router forwards the packet with the header in the set $S\textsuperscript{affected}$.
Since nodes of the binary tree store information of the router's identifier ($i$) and corresponding port ($p$) in the form of a tuple ($i,\;p$); therefore, during the tree traversal (as described for $S\textsuperscript{affected}$), Vercel collects tuples from the nodes labeled as atomic/supernet and includes those in the set $P\textsuperscript{affected}$.

The following theorem shows that it is possible to identify sets $S\textsuperscript{affected}$ and $P\textsuperscript{affected}$ in linear time, which justifies the fast verification time of Vercel (Section \ref{sec:evaluation}).

\textbf{Theorem 2}: The worst case time complexity to identify both the sets $S\textsuperscript{affected}$ and $P\textsuperscript{affected}$ is $\mathcal{O}(m)$, where $m$ is the number of atomic+iatomic headers in the network.


\textit{Proof of Theorem 2:}
Insertion of a rule requires finding affected headers and relevant router ports (Section \ref{sec:design}.B).
Insertion of a header (corresponding to a new rule) in the binary tree requires $\mathcal{O}(m)$ time for traversing a path length of $L$ (for $L$ bits header) from the root to a node $o_j$ and visiting all child nodes of $o_j$ in order to create iatomic nodes in the tree (worst case scenario).
While traversing the tree, Vercel finds all overlapping rules in a router, creates the set $S\textsuperscript{affected}$ by appending the identifier of the leaves and identifies relevant router ports $P\textsuperscript{affected}$ by collecting the $(i,\;p)$ tuples. This tree traversal a) from the root to node $o_j$ and b) subtrees of $o_j$ can be done with the time complexity of $\mathcal{O}(m)$.
Therefore, after a rule insertion, the time complexity to create the sets  $S\textsuperscript{affected}$ and $P\textsuperscript{affected}$ is $\mathcal{O}(m)$.
A similar argument holds for the rule deletion that has a time complexity of $\mathcal{O}(m)$.

\subsection{Creation of subspaces for relevant ports}\label{subsec:subspaces}
After obtaining sets $S\textsuperscript{affected}$ and $P\textsuperscript{affected}$, Vercel represents atomic+iatomic headers from the set $S\textsuperscript{affected}$ in a vector space of $m$-dimensions.
For that, Vercel computes the number of headers in the set $S\textsuperscript{affected}$ i.e., $m=|S\textsuperscript{affected}|$.
Thereafter, Vercel defines an $m$-dimensional space by creating $m$ orthogonal vectors and uniquely maps these vectors to $m$ atomic+iatomic headers in the set $S\textsuperscript{affected}$.
Orthogonality is essential to model the packet forwarding at routers with least squares (Section \ref{subsec:leastsq}).
The set of orthogonal vectors can be generated using the Gram-Schmidt \cite{gram} process (in which we take any vector, and then using that as a reference, make all other vectors orthogonal, one vector at a time). 
Later, we shall do away with requirement of orthogonal vectors while achieving linear time reachability computations (Section \ref{sec:reduce_time}).


After making an $m$-dimensional space, Vercel makes subspaces for each port. A port's subspace includes packets with atomic and iatomic headers that a router sends to that port. To make these subspaces, Vercel goes through each ($i,\;p$) tuple in the set $P\textsuperscript{affected}$. For each tuple, it picks packets with atomic and iatomic headers from the set $S\textsuperscript{affected}$ that router $i$ sends through output port $p$. 
If router $i$ sends packets with a unique set of $1\leq n\leq m$ headers through port $p$, Vercel creates a subspace for the ($i,\;p$) tuple. 
To do so, Vercel picks $n$ orthogonal vectors and stores them in a matrix $\mathcal{A}_i^p$ of size $(m,\;n)$. In addition to $\mathcal{A}_i^p$, Vercel makes a vector $b_{init}$ with $m$ dimensions. This vector shows the packet headers being evaluated for reachability between a source and destination.

\subsection{Least squares to model the packet processing at a router}\label{subsec:leastsq}

After creating subspaces, Vercel utilizes ports in the set $P\textsuperscript{affected}$ to traverse different paths between the source and destination for evaluating reachability.
Assume source $S$, destination $D$ and an interim router $i$, denoted by path: $S,...,i-1, i, i+1,...,D$.
Also, assume that Vercel has computed packets (with headers in $S\textsuperscript{affected}$) that are reachable up to router $i$ from $S$.
These packet headers are represented by vector $b_{i-1}$.
Now, Vercel aims to identify those headers in $b_{i-1}$ that router $i$ forwards to $i+1$ through one of its output ports $p$.
Algebraically, this situation can be achieved by first solving $\mathcal{A}_i^px_i^p=b_{i-1}$ to obtain $\hat{x}_i^p$.
Thereafter, with $\hat{x}_i^p$, Vercel obtains an orthogonal projection of $b_{i-1}$ onto the column space of $\mathcal{A}_i^p$ (denoted as $C(\mathcal{A}_i^p)$). 
The column space of a matrix denotes those points that can be obtained through a linear combination of columns of the matrix.
The efficient way to solve $\mathcal{A}_i^px_i^p=b_{i-1}$ is least squares, which always guarantees to provide a solution in real-time. 
While obtaining the projection of $b_{i-1}$ in $C(\mathcal{A}_i^p)$ using least squares, there can be three possible outcomes:

\textit{Case 1}: 
This is the dominant case in which router $i$ forwards \textit{some} of the received packets along an output port $p$. 
This is because the forwarding table has some of the matching rules for all incoming packets.
Mathematically, $b_{i-1}$ is partially both in the column space of $\mathcal{A}_i^p$ and the null space of $(\mathcal{A}_i^p)^T$.
The null space of a matrix denotes points that are orthogonal to all rows of the matrix.
In other words, $b_{i-1}$ is a linear combination of basis vectors from $C(\mathcal{A}_i^p)$ and $N((\mathcal{A}_i^p)^T))$.
As a consequence, there exists no solution to the linear equations $\mathcal{A}_i^px_i^p=b_{i-1}$.
Therefore, we utilize least squares to obtain the best approximate solution ($\hat{x}_i^p$).
Subsequently, $\hat{x}$ is used to obtain a projection point $b_i=\mathcal{A}_i^p\hat{x}_i^p$.
Now, the projected point $b_i$ (present in $C(\mathcal{A}_i^p)$), represents a subset of packet headers received at router $i$ (from the previous router $i-1$) which are being forwarded along output port $p$.
For an example of this case, refer to Fig. \ref{fig:projection_matrices}, where nodes $Y$ and $U$ are forwarding some of the prefixes in $b_i$.

\textit{Case 2}: 
In this case, router $i$ forwards all the received packets along its output port $p$.
Equivalently, $b_{i-1}$ is present in the column space of $\mathcal{A}_i^p$.
In this case, the method of least squares returns an exact solution $x_i^p$.
The projected point $b_i=\mathcal{A}_i^px_i^p$ contains information on all atomic+iatomic headers present in $b_{i-1}$.
Therefore, with least squares, Vercel models packet forwarding for all headers in $b_{i-1}$ to the next router $i+1$ along the port $p$ at router $i$.

\textit{Case 3}:
In this case, router $i$ blocks all the received packets along its output port $p$, implying $b_{i-1}$ lies in the null space of $(\mathcal{A}_i^p)^T$; therefore, least squares returns a solution $\hat{x}_i^p=0$.
The projection point is $b_i=\mathcal{A}_i^p\hat{x}_i^p=0$.
The projection point $b_i=0$ specifies that router $i$ blocks all packets with atomic+iatomic headers represented by $b_{i-1}$ at output port $p$. 

\subsection{Determining the reachable set of packet headers from the vector representation}\label{subsec:decode}
If there exist multiple paths between a source and destination, then Vercel would compute reachability along each path and store the reachable set of packet headers at the destination as $b_{reachable}= \sum_{d=1}^{e} b_{d}$, where $b_{d}$ is the set of reachable headers at the destination node along path $d$ and $e$ is the total number of paths created with ports in $P\textsuperscript{affected}$.
After traversing all the paths, Vercel labels an atomic+iatomic header (in the set $S\textsuperscript{affected}$) as reachable if it obtains a non-zero dot product between the vector representation of the header and $b_{reachable}$. 

\section{Vercel Optimization: Reducing Verification Time}\label{sec:reduce_time}
As discussed, Vercel models the packet forwarding behavior along each output port of a router by projecting a point in the column space of the matrix corresponding to the port.
The projection is generally obtained by solving normal equations $b_i=\mathcal{A}(\mathcal{A}^T\mathcal{A})^{-1}b_{i-1}$ (for simplicity, we have dropped the subscript and superscript for matrix $\mathcal{A}$ created for port $p$ of router $i$).
However, by selecting orthogonal vectors for representing atomic+iatomic headers, we can obtain the same projection point as $b_i=\mathcal{A}^p_i(\mathcal{A}^p_i)^Tb_{i-1}$.
Further, with standard basis vectors, the matrix product $\mathcal{A}^p_i(\mathcal{A}^p_i)^T \in \mathbb{R}^{m\times m}$ results in a diagonal matrix with entries 0 and 1.
The $j\textsuperscript{th}$ diagonal entry will be ``1", if $j$\textsuperscript{th} row of $\mathcal{A}^p_i$ is non-zero.
Therefore, vector $b_i$ is an element-wise product between the diagonal elements of $\mathcal{A}^p_i(\mathcal{A}^p_i)^T$ and $b_{i-1}$. 
We can efficiently obtain the diagonal elements of $\mathcal{A}^p_i(\mathcal{A}^p_i)^T$ by first creating $v_i^p=0$ in $m$-dimensional space and then initializing the $j\textsuperscript{th}$ entry of $v_i^p$ to 1 if router $i$ forwards $j\textsuperscript{th}$ header (in the set $S\textsuperscript{affected}$) to its output port $p$.
After initialization, the forwarding vector $v_i^p$ denotes atomic+iatomic headers (in the set $S\textsuperscript{affected}$) that router $i$ forwards to its output port $p$.
Subsequently, we can efficiently compute the projection point $b_i$ as $b_i = v_i^p \otimes b_{i-1}$, where $\otimes$ is the Hadamard product between two vectors.
Note that the projection point $b_i$ can now be computed efficiently in linear time.

\textbf{Theorem 3}\label{subsec:improve_complexity}:
The time complexity of modeling the forwarding behavior of a router along a port $p$ (i.e., obtaining a projection point $b_i\in \mathbb{R}^m$ from $b_{i-1}\in \mathbb{R}^m$) is $\mathcal{O}(m)$, where $m$ is the number of atomic+iatomic headers in set $S\textsuperscript{affected}$ with headers represented by the standard basis vectors.

\textit{Proof}: The proof is provided in Appendix I.

As described in Section \ref{sec:design}, at the destination, Vercel utilizes vector $b_{reachable}$ to represent those packet headers, which are reachable from the source.
The following theorem shows that by using the standard basis for representing atomic+iatomic headers, it is possible to efficiently recover the reachable packet headers from the vector $b_{reachable}$ in linear time.

\textbf{Theorem 4}:
The worst-case complexity of determining the reachable set of atomic+iatomic headers from $b_{reachable}\in \mathbb{R}^m$ is $\mathcal{O}(m)$.

\textit{Proof}: The proof is provided in Appendix II.

We have established that there exists a linear time solution: a) to model packet processing at a router using vector algebra; b) to determine the reachable set of packet headers from the vector $b_{reachable}$.
The following theorem now presents an efficient linear-time solution to compute the reachable set of headers along a path from source to destination.
\textit{This theorem is an improvement over the complexity achieved by state-of-art Deltanet, which is amortized linear in the number of affected headers and logarithmic in the number of overlapping rules present in a router}. 
In the worst case, the number of elements in the set $S\textsuperscript{affected}$ can equal the rules in the network.
However, with experiments (Section \ref{sec:evaluation}), we show that in all practical scenarios, the number of atomic+iatomic headers in the set $S\textsuperscript{affected}$ is much smaller than the total number of rules across all the devices.
Vercel implements the linear time approach suggested by the following theorem to achieve fast verification time on different networking scenarios.

\textbf{Theorem 5}:
In the worst case, the time complexity of reachability computation along a path using Vercel is $\mathcal{O}(m)$, where $m$ is the number of rules in the network.

\textit{Proof}: The proof is provided in Appendix III.

Although the above theorem provides time complexity for a single path, in practice, Vercel examines all possible paths (derived from $P\textsuperscript{affected}$) while checking reachability.
Next, we show memory efficiency of Vercel in terms of the size and number of forwarding vectors needed to check reachability between a source and destination.

\textbf{Theorem 6}:
After a rule insertion/deletion, the space complexity of Vercel is $\mathcal{O}(|S\textsuperscript{affected}|*|P\textsuperscript{affected}|)$ by using the standard basis to represent atomic+iatomic headers and forwarding vectors $v_i^p$).

\textit{Proof}: The proof is provided in Appendix IV.

\section{Expressive Vercel Features}\label{sec:beyond}

We now extend Vercel to incorporate reachability queries involving header transformations and packet filters, as well as for identifying forwarding loops and blackholes.

\subsection{Packet transformation}\label{subsec:IPtransform}
At some middleboxes in the network, forwarding tables also contain rules that transform packet headers,
thereby making reachability computation complex.
In general, \textit{header transformation} (denoted by function $f^{tr}$) is a set of rules containing the header ``match" and ``transformed" fields. 
To model header transformations, Vercel extracts transformation rules from routers and inserts the headers (specified in the ``match" and ``transformed" fields) in the binary tree (as discussed in Section \ref{subsec:bintree}, \ref{subsec:nodesbintree}).
Thereafter, Vercel creates a transformation matrix $T_i\in \mathbb{R}^{m\times m}$, for each router $i$ performing header transformations.
Here, $m$ denotes the total number of atomic+iatomic headers in the network.
Vercel initializes all entries of matrix $T$ to zero, representing the absence of header transformation.
Thereafter, a few entries in the matrix $T$ are updated to 1, based on the transformation rules.
An entry $T^{(j,k)}=1$ of the matrix specifies that an atomic/iatomic header with identifier $k$ is transformed to another atomic/iatomic header with identifier $j$.
While solving for reachability, Vercel first computes $t_i = H(T_i b_{i-1})$ to transform the headers and then solves $\mathcal{A}_i^px_i^p = t_i$ to model forwarding at router $i$ (where $H$ represents the unit step function).
The vector $t_i$ represents atomic+iatomic headers reachable up to the router $i$ and then transformed using $T_i$. 

\textit{Example of Transformation Matrix:}
Consider the headers in the forwarding table of router $U$ (Figure \ref{fig:toynet_new}) and a transformation function (denoted as $f^{tr}_U$) containing a single rule $f^{tr}_U:01/2\to 00/2$.
To create the transformation matrix at router $U$, i.e., $T_U$, Vercel requires information of atomic+iatomic headers ahead in time.
In this example, the atomic and iatomic headers are $\{a1, q2, a3, a4\}$.
The matched header 01/2 ($a3$) is itself atomic while transformed header 00/2 can be represented by a combination of an atomic 000/3 ($a1$) and an iatomic 001/3 ($q2$) headers.
Now, $T_U$ is a (4, 4) matrix populated as follows.

{\let\quad\thinspace
\vspace*{-10pt}
\begin{equation*}
T_U =\;\; \scriptsize \bordermatrix{
      &a1 &q2 &a3 &a4   \cr
	a1& 1 & 0 & 1 & 0   \cr
    q2& 0 & 1 & 1 & 0   \cr
    a3& 0 & 0 & 0 & 0   \cr
    a4& 0 & 0 & 0 & 1   \cr
 },\;
\vspace*{3pt}
\end{equation*}
}

The transformation matrix $T_U$ is initialized with ``1"s along its diagonal, except for the index (3, 3).
To represent the transformation of header $a3$ to ($a1,\; q2$), $T_U$ is initialized with ``1"s at indices (1, 3) and (2, 3).

\subsection{Packet filtering}\label{sec:packet_filter}

In a network, devices such as firewalls and routers contain an access control list (ACL) to restrict the access of packets.
A rule in an ACL is a filtering function $f^{acl}$ that maps a set of packet headers to a set of \{permit, deny\} actions.
To incorporate packet filtering at a router/firewall $i$, Vercel extracts ACL rules from $i$ and arranges the header fields from ACL rules in the existing binary tree.
The process of adding ACL rules to the binary tree is similar to that of adding forwarding rules (Section \ref{subsec:bintree}), though with different actions.
Thereafter, to model packet filtering for all the atomic+iatomic headers in the affected set (Section \ref{subsec:affectedbintree}), Vercel creates an $m$-dimensional filtering vector $g_i$ with binary values (0/1).
A non-zero entry at the $j$\textsuperscript{th} index of the vector $g_i$ implies that an ACL rule allows packets with atomic/iatomic header $s_j$ to pass through router $i$.
After obtaining the filtering vector $g_i$, Vercel performs filtering at router $i$ as $f_i=g_i\otimes b_{i-1}$, where $\otimes$ represents the Hadamard product between vectors.
An index $j$ with the non-zero entry in the filtered vector $f_i$ implies that received packets with atomic/iatomic header $s_j$ are permitted to pass through router $i$.
After filtering, Vercel models packet forwarding along port $p$ at router $i$ by solving  $\mathcal{A}_i^px_i=f_i$ with least squares.

\subsection{Identifying forwarding loop}\label{subsec:loops}

To detect a forwarding loop, consider an intermediate router $i$, along the path $S$,...,$i$-$1$, $i$, $i$+$1$,...,$D$. 
Router $i$ receives packets with headers in vector $b_{i-1}$ from router $i$-$1$. 
Thereafter, Vercel models packet forwarding at router $i$ along a port $p$ using $\mathcal{A}_i^px_i^p=b_{i-1}$ and obtains a projection point $b_i\neq 0$.
After obtaining $b_i$, Vercel checks whether the next router $i+1$ (connected through port $p$ at router $i$) has already been traversed by the received packets.
If the identifier of the next router $i+1$ is already traversed, then Vercel confirms that the rule update will trigger a loop.
To compute which packet headers are in a loop, Vercel searches for the indices with non-zero entries in $b_i$.

\subsection{Identifying blackholes}\label{subsec:blackholes}

Blackholes represent a network state in which a router receives a packet, but its forwarding table does not contain a corresponding rule.
To identify a blackhole at router $i$, Vercel implements the following steps.
(1) Vercel models the forwarding behavior of router $i$ based on the rules in its forwarding table. 
Since we have $v_i^p$ for all the affected ports of $i$ (Section \ref{subsec:affectedbintree}, \ref{sec:reduce_time}), Vercel performs element-wise logical OR between all $v_i^p$ (denoting atomic+iatomic headers that router $i$ forwards to its output port $p$) to get a unified forwarding vector $v_i$ corresponding to router $i$.
Intuitively, $v_i$ represents packets with atomic+iatomic headers that router $i$ forwards to its neighbors.
(2) Vercel computes $v_i\otimes b_{i-1}$ to obtain a projection point $b_i$.
Vector $b_i$ represents those atomic+iatomic headers in $b_{i-1}$ that router $i$ forwards to its neighbors. 
(3) To detect a blackhole at router $i$, Vercel computes $c_i = b_{i-1}\oplus b_{i}$ (where $\oplus$ denotes the element-wise logical XOR).
The indices with a non-zero entry in vector $c_i$ represent packets (with atomic+iatomic headers) received at router $i$ that do not get any match in the forwarding table.

\subsection{ Recommendations from projection errors}\label{subsec:pathreco}

The approximate solution $\hat{x}$ obtained with least squares provides a projection point $\mathcal{A}\hat{x}$.
We can measure the difference between two points $\mathcal{A}\hat{x}$ and $b$ to obtain the \textit{projection error}, whose absolute value for computation of reachability has no implication.
However, the projection error ($\mathcal{A}\hat{x}-b$) is a useful metric to provide intuition into packet processing at a router as well as network-wide metric computations.
While the quantum of projection error ($\mathcal{A}\hat{x}-b$) has no bearing on reachability, it  turns out that the cumulative error vector could be useful in identifying path correctness. 
In short, the cumulative $l2$ norm of projection errors across a path can reflect the misconfigurations along the path. 

When a router receives a packet at a port, then there must exist a rule that forwards the packet to another port. 
However, if no rule exists that matches the address field of this packet, then taking $\mathcal{A}\hat{x}-b$  contributes towards a higher projection error.
If the configuration is not along the shortest path, then even if the projection error at a router is low, the accumulated projection error along the path adds up and becomes high.

Hence, one use of projection error computation is to find if the configuration is on the shortest path, which gives a direct measure of the data plane correctness. 
To achieve this, the controller computes the $l2$ norm of the error at each node and adds the errors along all possible paths obtained through router configurations.
If the configurations are correct, the shortest path should have the least total error. If this is not the case, Vercel adjusts the configurations based on the errors at intermediate nodes to ensure they result in the shortest path.

For more insight into this, suppose a provider wants to ensure that all the flows belonging to say a VR application must be configured always on the shortest path.
To this end, Vercel initializes vector $b_0$ with the IP addresses corresponding to these flows and runs a reachability check. 
While doing so, it computes the cumulative error along each path to the destination. 
In case the minimum cumulative error is not obtained from the shortest path, then Vercel identifies that most of the configuration is on longer paths.
Now, by observing each intermediate node's projection error, Vercel identifies which flows do not have a matching rule along the shortest path. 
Finally, these missing rules are configured at the identified nodes. 
To ensure that the non-shortest path is not chosen, the corresponding rules are first identified and then deleted from the nodes along the non-shortest paths by using projection errors. 

\textit{Example:} To understand how projection errors can be used prolifically, revert back to Figure \ref{fig:projection_example}, we also observe that router $U$ drops some packets due to the absence of a rule that will process header $001/3$.
This is observed by computing the projection error between vectors $b_Y=$ [1, 1, 0] (headers received at $U$) and $b_U=$ [0, 1, 0] (headers forwarded by $U$ along port 0).
The projection error ($b_Y-b_U$) along port 0 at router $U$ is [1, 0, 0], and this indicates that packets with header 001/3 are not forwarded along the output port 0 at router $U$.
However, as per the shortest path configuration, router $U$ should forward packets with header 001/3 along port 0 (in which case the projection error is zero).
Therefore, the error vector helps in identifying possible misconfigurations.

A second use of the projection error is for automatic table population amidst no reachability. 
If reachability does not exist between two routers, then the cumulative projection error will be very large along all the paths.
A path with the lowest cumulative \textit{l2} norm indicates that there are already a large number of relevant rules installed at the routers, and we just have to add a few more to establish reachability.
On that path, for the nodes which have high projection error (compared to other nodes), Vercel recommends the addition of table entries such that the projection error at those nodes reduces. 
The table entries (prefix, port), now facilitate the router at which the entry is added to route the packet to the destination, leading to reachability.
Through this iterative process, the entries can be updated to result in the shortest path. 
In this way, by using a simple script, Vercel can automatically establish reachability while achieving the shortest path routing for the source-destination pairs of interest.

\section{Evaluation}\label{sec:evaluation}
We evaluate Vercel as a verification tool on various datasets and compare it with other approaches. 

\subsection{Implementation and dataset details}\label{subsec:dataset}

Vercel is implemented as $\sim$2000 lines of single-threaded Python 3.7 code with dependency on Numpy \cite{Numpy} (for matrix computations) and Networkx \cite{Networkx} (for creating a network graph).
We evaluate the performance of Vercel on a single core of an Intel Core i7 CPU at 3.6GHz clock and 64GB RAM.

To extensively evaluate the performance of Vercel, we have considered network topologies ranging from campus networks to large enterprise/provider networks with millions of rules (up to 124.7 million).
Shown in Table \ref{tab:dataset} are the datasets that we consider for Vercel's evaluation.
The Stanford and Berkeley datasets serve as benchmarks for evaluating network invariants \cite{stanford-datset, deltanet-datset}. 
Further, we also consider datasets RF 1755, RF 6461, RF 3257, and INET consisting of rules from the autonomous systems derived from the Rocketfuel Project \cite{Rocketfuel}. 
These datasets represent dense network topologies and contain millions of forwarding rules, suitable for evaluatng scalability of a verification technique. 
To evaluate Vercel's performance on a service provider network, we consider datasets Airtel 1, Airtel 2 used by \cite{delta-net}, which are available at \cite{deltanet-datset}.
To expand to very large providers, we created two synthetic datasets -- Simnet1 and Simnet2, containing 1000 and 2000 nodes with 100K edges in each network.
We assign IP addresses on these two networks based on the mask length distribution extracted from the Rocketfuel project.
Finally, we populate forwarding rules on Simnet1 and Simnet2 with shortest path routing.
We also point out that Vercel has been implemented in conjunction with controllers and switches such as ONOS and OVS, thus showing industry-centric compatibility.
Table \ref{tab:dataset} provides the specifics of the datasets such as nodes, edges, number of rules in the network, number of updates (insertion/deletion).
To compute the verification time of Vercel, we first load 90\% of the forwarding rules from the dataset, then populate the binary tree and then perform real-time updates by randomly selecting the remaining rules for insertion and deletion at router forwarding tables. 
We consider dynamic updates by first performing rule insertion in the forwarding tables (which comprises half of the rule updates shown in the sixth column of Table \ref{tab:dataset}), followed by rule deletion. 

\begin{table}[t]
\caption{Details of the datasets.}
\label{tab:dataset}
\centering
\setlength{\tabcolsep}{3pt} 
\renewcommand{\arraystretch}{1.2} 
\begin{scriptsize}
 \begin{tabular}{|m{1cm}|m{0.7cm}|m{0.8cm}|m{0.9cm}|m{1.4cm}|m{0.9cm}|m{0.9cm}|}
 \hline
 \rowcolor{lightgray}
 Dataset & Nodes & Edges & Average degree of nodes & Number of rules & Number of rule updates & Average table size \\
 \hline
 Airtel1 & 16 & 26 & 3.25 & 3.81 $\times$ 10\textsuperscript{4} & 2 $\times$ 10\textsuperscript{5} & 2381\\
 \hline
 Airtel2 & 16 & 26 & 3.25 & 3.81 $\times$ 10\textsuperscript{4} & 2 $\times$ 10\textsuperscript{5} & 2381\\
 \hline
 Stanford & 16 & 74 & 9.25 & 7.3 $\times$ 10\textsuperscript{5} & 1 $\times$ 10\textsuperscript{4} & 45570\\
 \hline
 Berkeley & 23 & 252 & 21.91 & 12.81 $\times$ 10\textsuperscript{6} & 1 $\times$ 10\textsuperscript{4} & 557300\\
 \hline
 RF 1755 & 87 & 2308 & 53.06 & 33.73 $\times$ 10\textsuperscript{6} & 1 $\times$ 10\textsuperscript{4} & 387734\\
 \hline
 RF 6461 & 138 & 8140 & 117.97 & 75.01 $\times$ 10\textsuperscript{6} & 1 $\times$ 10\textsuperscript{4} & 543519\\
 \hline
 RF 3257 & 161 & 9432 & 117.17 & 74.49 $\times$ 10\textsuperscript{6} & 1 $\times$ 10\textsuperscript{4} & 422688\\
 \hline
 INET & 314 & 40770 & 259.68 & 124.7 $\times$ 10\textsuperscript{6} & 1 $\times$ 10\textsuperscript{4} & 395979\\
 \hline
 Simnet1 & 1000 & 100000 & 200 & 49.95 $\times$ 10\textsuperscript{6} & 1 $\times$ 10\textsuperscript{4} & 49950\\
 \hline
 Simnet2 & 2000 & 100000 & 100 & 99.95 $\times$ 10\textsuperscript{6} & 1 $\times$ 10\textsuperscript{4} & 49975\\
 \hline
 \end{tabular}
\end{scriptsize}
\end{table}

\subsection{Verification time per rule update}\label{subsec:verifybasic}
\begin{figure}
    \centering
    \includegraphics[width=0.9\linewidth]{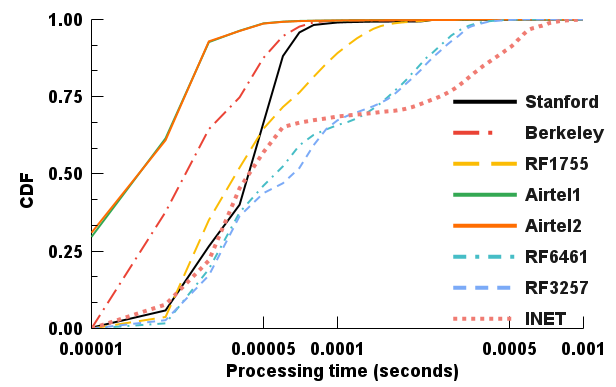}
    \caption{CDF of the verification time of Vercel on 8 datasets \cite{deltanet-datset, stanford-datset} during a rule update.}
    \label{fig:cdf_reachability}
    \vspace*{-15pt}
\end{figure}

The most important metric to evaluate a network verification scheme is \ul{verification time} \cite{veriflow, delta-net, apkeep, netplumber}. 
Figure \ref{fig:cdf_reachability} shows the cumulative distribution function (CDF) of verification time achieved by Vercel applied to the datasets \cite{deltanet-datset, stanford-datset}. 
From Figure \ref{fig:cdf_reachability}, we observe that Vercel processes at least 70\% of the updates within $100\mu s$ on all the datasets.
Even if we consider 90\% of the updates, verification time of Vercel remains under $500\mu s$. 
These results imply that Vercel achieves sub-millisecond verification time, even on large networks. 

\begin{table*}[t]
\caption{Rule verification time of Vercel for checking reachability.}
\label{tab:performance}
\centering
\setlength{\tabcolsep}{5pt} 
\begin{scriptsize}
 \begin{tabular}{|m{2.7cm}|m{1cm}|m{1cm}|m{1cm}|m{1cm}|m{1cm}|m{1cm}|m{1cm}|m{1cm}|m{1cm}|m{1cm}|} 
 \hline
 \rowcolor{lightgray}
 Dataset & Airtel1 & Airtel2 & Stanford & Berkeley & RF1755 & RF6461 & RF3257 & INET & Simnet1 & Simnet2 \\
 \hline
 \# atomic+iatomic headers & 1404 & 1404 & 335225 & 747854 & 902740 & 902740 & 902740 & 629954 & 77437 & 77836 \\ 
\hhline{|=|=|=|=|=|=|=|=|=|=|=|}
 Median (in $\mu s$) & 26 & 26 &	53 & 52 & 32 & 66 & 77 & 53 & 112 & 194 \\
 \hline
 Mean (in $\mu s$) & 32 & 32 & 53 & 55 & 40 & 111 & 116 & 162 & 232 & 428\\
 \hline
 Percentage \textless $250 \mu s$ & 99.76\% & 99.76\% & 99.75\% & 99.8\% & 99.72\% & 88.72\% & 87.07\% & 73.9\% & 70.9\% & 54.58\% \\
 \hline
 \end{tabular}
 \end{scriptsize}
\end{table*}

Table \ref{tab:performance} provides verification time of Vercel on 10 datasets.
The first row of Table \ref{tab:performance} shows the total number of atomic+ iatomic headers in each network. 
Note that the ratio of atomic+iatomic headers to the number of forwarding rules on the INET dataset (containing the largest number of rules across all datasets) is as small as $\sim$0.005, hence Vercel exploits the overlap among forwarding rules and represents the header space with a small number of non-overlapping atomic+iatomic headers, thereby reducing the number of headers to be processed. 
The second and third rows of Table \ref{tab:performance} show the median and mean of the verification time for Vercel on different datasets.
These results show that the median and mean verification time of Vercel remains bounded to within $77\mu s$ and $162\mu s$ on all but the two synthetic datasets.
In the case of the INET dataset, the mean increases to $162\mu s$ as the average number of neighbors per router is $\sim$260 (which is significantly higher than all other datasets).
This increase in nodal degree results in the traversal of more paths in the network, increasing verification time.
We also observe the mean and median of verification time of Vercel on two synthetic networks is up to $428\mu s$ and $194\mu s$.
Since these networks contain a large number of (2000) nodes and (100K) edges, therefore reachability analysis requires traversing longer paths between source and destination nodes.
Accordingly, an increase in the path length increases verification time.

Another metric of interest (benchmarked by \cite{delta-net, apkeep}) is -- how many updates are verified in less than $250\mu s$ (third row in Table \ref{tab:performance}).
Vercel verifies $\sim$74\% of the updates within $250\mu s$, on a large network with $\sim$124.7M rules.
For a network with $2000$ nodes, Vercel manages to verify $\sim$55\% of the rules within $250\mu s$.
The primary reason for this fast verification time is the quick identification of the affected atomic+iatomic headers and verifying multiple headers together in vector space using least squares.

\subsection{Identifying loops and blackhole}\label{subsec:results_loops}
Shown in Table \ref{tab:performance2} (rows labeled with RL and RB) is the time for determining reachability in the presence of loops and in another case with blackholes.
In these two cases, the mean verification time of Vercel is less than 90 and 98 $\mu$s on all the datasets.
Verifying loops require a slightly longer time because Vercel performs extra checks for re-traversal along a path and stores a non-zero vector $b_i$ to identify the headers in the loop (Section \ref{sec:beyond}). 
For blackhole detection, Vercel performs some extra steps (element-wise logical OR and XOR, see Section \ref{sec:beyond}) along with a reachability check, which slightly increases the time for verification.

\begin{table}[h]
\caption{Vercel's reachability evaluation in the presence of loops (RL), blackhole (RB), packet transformations (RT), packet filtering (RF) and routing policy (RP) analysis.}
\label{tab:performance2}
\centering
\setlength{\tabcolsep}{5pt} 
\begin{scriptsize}
 \begin{tabular}{|m{1.8cm}|m{1cm}|m{1cm}|m{0.9cm}|m{0.7cm}|m{0.7cm}|} 
 \hline
 \rowcolor{lightgray}
  Dataset & Stanford & Berkeley & RF 1755 & Airtel 1 & Airtel 2 \\
 \hline
 Median (RL) & 76$\mu s$ & 53$\mu s$ & 88$\mu s$ & 33$\mu s$ & 32$\mu s$ \\
 \hline
 Mean (RL) & 79$\mu s$ & 56$\mu s$ & 92$\mu s$ & 63$\mu s$ & 57$\mu s$ \\
 \hhline{|=|=|=|=|=|=|}
  Median (RB) & 96$\mu s$ & 53$\mu s$ & 91$\mu s$ & 34$\mu s$ & 34$\mu s$ \\
 \hline
 Mean (RB) & 98$\mu s$ & 59$\mu s$ & 97$\mu s$ & 64$\mu s$ & 59$\mu s$ \\
 \hhline{|=|=|=|=|=|=|}
 Median (RT) & 884$\mu s$ & 417$\mu s$ & 469$\mu s$ & 36$\mu s$ & 37$\mu s$ \\
 \hline
 Mean (RT) & 918$\mu s$ & 349$\mu s$ & 440$\mu s$ & 285$\mu s$ & 229$\mu s$ \\
 \hhline{|=|=|=|=|=|=|}
 Median (RF) & 75$\mu s$ & 53$\mu s$ & 87$\mu s$ & 32$\mu s$ & 31$\mu s$ \\
 \hline
 Mean (RF) & 77$\mu s$ & 56$\mu s$ & 91$\mu s$ & 49$\mu s$ & 45$\mu s$ \\
\hhline{|=|=|=|=|=|=|}
 Median (RP) & 82$\mu s$ & 62$\mu s$ & 88$\mu s$ & 34$\mu s$ & 33$\mu s$ \\
 \hline
 Mean (RP) & 85$\mu s$ & 60$\mu s$ & 96$\mu s$ & 38$\mu s$ & 36$\mu s$ \\
 \hline
 \end{tabular}
 \vspace*{-8pt}
 \end{scriptsize}
\end{table}

\subsection{Packet filtering}\label{subsec:results_filter}

Since none of the existing datasets provide ACL rules, we created ACL rules by mapping the destination IP addresses (in the existing datasets) to filtering actions (permit/deny).
Table \ref{tab:performance2} shows that the mean and median verification time in the presence of ACL rules (represented as rows labeled with RF for Vercel) is less than 91 and 87 $\mu$s on all the datasets that we considered.

\subsection{Packet header transformation}\label{subsec:results_iptrans}
We define a function to transform an IP prefix into another randomly chosen IP prefix and store the transformation function in a matrix $T_i$, for each router $i$ (Section \ref{sec:beyond}).
In Table \ref{tab:performance2} (with the rows labeled as RT), we observe that the mean verification time of Vercel is less than 1 ms on all the datasets.
Due to the matrix-vector product $T_ib_{i-1}$, the mean verification time increases in the presence of transformation functions.
Note that the verification time of Vercel is comparable to APKeep \cite{apkeep}, in which the verification time is also bounded within 1 ms.
Note, Veriflow \cite{veriflow} and Deltanet \cite{delta-net} do not model packet transformations.

\subsection{Checking policy violations}\label{subsec:results_policy}
Policy-based routing (PBR) \cite{policy} is used by network administrators to define the routing behavior of packets in a network by overriding the underlying routing protocol. 
Vercel assigns a PBR label to those nodes of the binary tree whose corresponding headers are generated through PBR.
Thereafter, Vercel does not allow any non-PBR protocol to update the forwarding action (e.g., output port) at the labeled node in the binary tree. 
Subsequently, after a rule update, Vercel determines the affected headers and ports in the binary tree.
Finally, Vercel evaluates reachability and checks for the violation of routing policies.
In our evaluation, Vercel identifies whether, after a rule update, the traffic: a) violates a path length constraint; or, b) passes through a predetermined set of routers.
Table \ref{tab:performance2} (rows labeled with RP) shows that the median and mean verification time of Vercel for checking these two policies together is less than 88 and 96 $\mu s$.
With an additional consideration of routing policy, the verification time increases slightly.
The reason is that after evaluating reachability at the destination, Vercel has to iterate over the reachable paths to check for policy violations.

\subsection{Path recommendation, error rectification}\label{subsec:results_path_recom_recti}
\begin{figure}[t]
    \centering
    \includegraphics[width=0.8\linewidth]{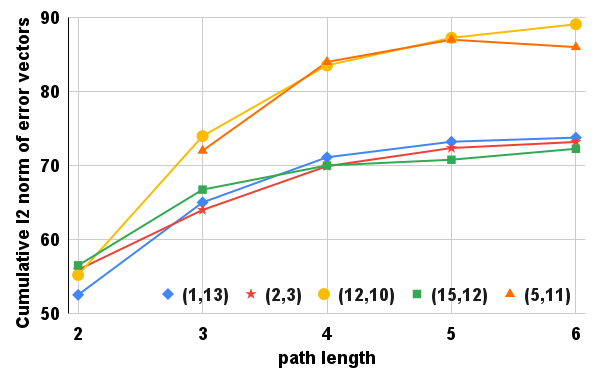}    \caption{Cumulative L2 norm of the projection error as a function of the path length for different source-destination pairs on Airtel1 dataset. }
    \label{fig:l2normVspathLen}
    \vspace*{-10pt}
\end{figure}

Through Vercel we determine the quality of different paths (for example when deploying ECMP) by utilizing the projection error accumulated over nodes along each path.
The cumulative $L2$-norm of the projection error (across all nodes along a path) quantifies the \textit{quality} of a path in terms of the number of misconfigured packet headers at routers on the path. 
Figure \ref{fig:l2normVspathLen} shows the cumulative $L2$ norm of the projection error as a function of  path length.
Since protocols configure most of the services on shortest paths, therefore we observe low projection error on such paths.
Generally, fewer services are configured on longer paths, which is reflected by an increase in the cumulative L2 norm of the projection error.
Therefore, we can use projection error as a metric to recommend the paths for service configurations. 

\begin{figure}[t]
    \centering
    \includegraphics[width=0.85\linewidth]{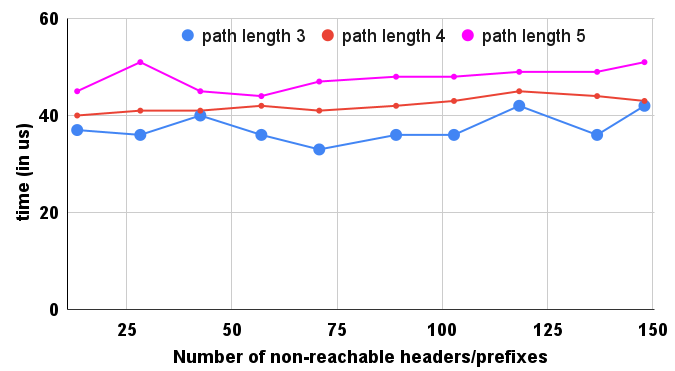}
    \caption{Computation time of Vercel for error rectification as a function of non-reachable headers and path length.}
    \label{fig:errorVstime}
    \vspace*{-10pt}
\end{figure}

As discussed in Section \ref{subsec:rect_prov_recomm}, error rectification is done by including $b$ in the column space of matrix $\mathcal{A}$ for all nodes along the path, where $b$ is in the null space of $\mathcal{A}^T$.
We created a synthetic network of 50 nodes and 200 edges to evaluate this approach and configured corresponding forwarding tables. 
Now, Vercel is applied to identify reachability between an arbitrary source and destination node. Assume, Vercel recognizes that there is no reachability, indicating that a fix is required. Computing this fix must be done fast. To show rectification computation, we varied the number of atomic/iatomic headers in the network and measured the time to rectify across differing path lengths. Results shown in Figure \ref{fig:errorVstime} allude to the time required to establish reachability.
A path length of five required more fixes than a shorter path length of three. However, even if we increase the network-wide number of atomic/iatomic headers, the fixing time does not grow linearly.
This is because Vercel leverages vectorization, where the same algebraic operation on different headers can be performed simultaneously.
From Figure \ref{fig:errorVstime}, it is evident that even on a longer path, Vercel can establish reachability within 50$\mu s$ irrespective of the number of atomic/iatomic headers.

\subsection{Comparison}\label{subsec:results_comparison}
We compare Vercel with the state-of-art real-time data plane verification tools such as AP Verifier \cite{atomic-predicate}, Veriflow \cite{veriflow}, NetPlumber \cite{netplumber}, Deltanet \cite{delta-net} and APKeep \cite{apkeep}.
Table \ref{tab:compare} compares the verification time of the various tools on the Stanford (campus) and Airtel (service provider) datasets.
Note that on a large dataset such as Stanford, the performance of Vercel is better than all other verification techniques except Deltanet. Note Deltanet is designed only to model forwarding rules.
In contrast, Vercel, APKeep, NetPlumber, and AP Verifier support data plane verification with a larger set of network functions such as filtering and transformation.
Table \ref{tab:compare} shows that, on the Stanford dataset, Vercel achieves mean verification time of 53 $\mu$s and its performance is summarized as:  $8\times$ over Veriflow, $164\times$ over NetPlumber, $1.7\times$ over APKeep, $36\times$ over AP Verifier.
On the Stanford dataset, Vercel processes 99.7\% of the rule updates within 250 $\mu$s.
Whereas, existing solutions such as NetPlumber, APVerifier, Veriflow and APKeep process 23.6\%, 13.3\%, 96.1\% and 96.4\% of the rule updates within 250 $\mu$s, respectively.
Similarly, on a small dataset of Airtel 1, Vercel achieves a significant improvement of $2.5\times$ over AP Verifier, $1.84\times$ over Veriflow, and $118\times$ over NetPlumber.
The reason for Vercel's speedup is the simultaneous processing of multiple atomic+iatomic headers at different ports.
In contrast, other approaches do not have the in-built capability of simultaneous processing and are comparatively slower.

\begin{table}[h]
\caption{Comparing verification time of different approaches on public datasets \cite{apkeep}. (TO = timeout)}
\label{tab:compare}
\setlength\tabcolsep{5pt} 
\centering
\begin{scriptsize}
 
 \begin{tabular}{|m{1.3cm}|m{0.6cm}|m{0.8cm}|m{0.6cm}|m{0.8cm}|m{0.6cm}|m{0.8cm}|}
 \hline
 \rowcolor{lightgray}
  Dataset & \multicolumn{2}{c|}{Stanford} & \multicolumn{2}{c|}{Airtel 1} & \multicolumn{2}{c|}{Airtel 2} \\
 \hline
 Metric & time $(\mu s)$ & \% \textless 250 $\mu s$ & time $(\mu s)$ & \% \textless 250 $\mu s$ & time $(\mu s)$ & \% \textless 250 $\mu s$\\
 \hhline{|=|=|=|=|=|=|=|}
 AP Verifier & 1953 & 13.3 & 80 & 91.3 & 135 & 77.4\\
 \hline
 Veriflow & 468 & 96.1 & 59 & 99.9 & 48 & 99.9\\
 \hline
 NetPlumber & 8700 & 23.6 & 3804 & 3.8 & TO & TO\\
 \hline
 Deltanet & 9 & 99.9 & 3 & 99.9 & 4 & 99.9\\
 \hline
 APKeep & 94 & 96.4 & 7 & 99.8 & 6 & 99.9\\
 \hline
 \textbf{Vercel} & \textbf{53} & \textbf{99.7} & \textbf{32} & \textbf{99.7} & \textbf{32} & \textbf{99.7}\\
 \hline
 \end{tabular}
 \vspace*{-10pt}
\end{scriptsize}
\end{table}

\subsection{What-if queries and Batch Processing}\label{subsec:results_whatif}
A network administrator can use ``what-if" queries to determine the fate of packets when a link goes down or model other scenarios.
After a link failure, the SDN controller/ routing protocol recomputes paths between different source-destination pairs.
Subsequently, the controller deletes existing rules corresponding to the failed links and inserts new rules (based on the updated paths) in the forwarding table of routers.
To simulate a link failure, we select a random link $(l)$ in the network and delete rules from the both the connected routers that forward packets over $l$.
As shown in the first row of Table \ref{tab:whatif}, a link failure can trigger deletion of 1000s of rules, therefore Vercel computes reachability by considering \textit{batches} of rules.
Table \ref{tab:whatif} shows that the mean verification time for checking reachability along with loops on the INET dataset for Vercel is around 534 ms when a link failure triggers $\sim$3060 rule deletions in the network.
In contrast, Deltanet and Veriflow require 2888 and 29117 ms.
This result shows that Vercel is up to 5$\times$ and 54$\times$ faster than Deltanet and Veriflow on INET dataset.
However, Deltanet performs better in 3 of the 5 datasets; though, the performance of Deltanet is dependent on network size and the number of rules.
\begin{table}[h]
\caption{Comparing different approaches for `what-if' queries for link failures. The second half of the table compares the memory utilization to results in \cite{delta-net}.}
\label{tab:whatif}
\setlength\tabcolsep{4pt} 
\centering
\begin{scriptsize}
 
 \begin{tabular}{|p{1.1cm}|wl{1cm}|wl{0.9cm}|wl{0.9cm}|wl{0.9cm}|wl{0.9cm}|wl{0.7cm}|}
 \hline
 \rowcolor{lightgray}
  \multicolumn{2}{|c|}{Dataset} & Berkeley & RF1755 & RF6461 & RF3257 & INET \\
 \hline
 \multicolumn{2}{|c|}{Avg. \# of updates} & 50863 & 14603 & 9216 & 7900 & 3060 \\
 \hline
 \multirow{3}{1cm}{Verification time (in ms) } & Veriflow & 3073 & 8100 & 17594 & 17645 & 29117 \\
 \cline{2-7}
 & Deltanet & 93 & 897 & 0.4 & 2.6 & 2888 \\
 \cline{2-7}
 & \textbf{Vercel} & \textbf{1083} & \textbf{603} & \textbf{766} & \textbf{741} & \textbf{534} \\
 \hline
 \hhline{|=|=|=|=|=|=|=|}
 \multirow{3}{1cm}{Memory (in MB) } & Veriflow & 1089 & 2713 & 5920 & 5882 & 9776\\
 \cline{2-7}
 & Deltanet & 6208 & 16937 & 39481 & 40716 & 63563 \\
 \cline{2-7}
 & \textbf{Vercel} & \textbf{2216} & \textbf{4207} & \textbf{5047} & \textbf{5529} & \textbf{14200} \\
 \hline
 \end{tabular}
 \vspace*{-15pt}
 \end{scriptsize}
\end{table}

\subsection{Memory utilization}\label{subsec:results_memory}
We now show the memory utilization of Vercel on the various datasets and compared to Veriflow and Deltanet.
Table \ref{tab:whatif} shows that Vercel is upto 7.8$\times$ memory efficient than Deltanet.
Deltanet is memory intensive due to the implementation of two data structures: 1) binary search tree to create atoms and 2) edge labeled graph to model the forwarding behavior of packets in a network.
As compared to Veriflow, Vercel consumes more memory because of the extra space required to store: a) (node, port) tuples at different nodes of the tree; b) node label in string format and; c) integer identifier at leaf nodes of the tree, but also leads to more expressive power and path recommendation.

\section{Related Work}\label{sec:related}

The initially proposed data plane verification techniques require a snapshot of the network state to perform queries such as reachability, loop detection and slice isolation \cite{static-reachability, hsa, anteater, atomic-predicate, al2010flowchecker, al2009network, fogel2015general, jeffrey2009model, lopes2015checking, libra, Jinjing, BUZZ}.

Xie et al. \cite{static-reachability} presented a technique for performing static reachability analysis of large-scale IP networks. The authors modelled routing protocols, packet filters and packet transformations using a formal framework. 
Their reachability check algorithm can be used to efficiently determine the reachability in IP networks and also capture select what-if scenarios. 

{
Mai et al. \cite{anteater} proposed a tool called Anteater that enables operators to debug their data plane by identifying bugs in forwarding behavior. Anteater works by comparing the expected behavior of network packets with the actual behavior observed on the network, using a set of customizable test cases. 
Anteater converts high-level network invariants into boolean satisfiability problems (SAT) and solves these using SAT solvers.
In case, the network state is found to violate the invariants, Anteater provides a counter-example that helps in tracking the root cause. This tool mainly focused on identifying forwarding loops, packet loss, inconsistencies that emerged through dataplane misconfiguration.


Header space analysis (HSA) \cite{hsa} is a key approach for performing static analysis of network configuration. Header transformation -- a critical step in HSA involves analyzing the effects of network policies on packet headers. Authors describe several types of header transformations, such as port mapping, address rewriting, and develop techniques for analyzing their effects on header spaces. Specifically, HSA represents a $L$-bit packet header in $L$-dimensional space with reachability checks translated to algebraic operations over $L$-dimensional hypercubes. The limitation of HSA is that it can be computationally expensive, which makes it unsuitable, especially for large and complex networks.

Snapshot-based techniques are designed to identify problems after they occur. However, in order to mitigate their impact, it is necessary to detect events before they cause problems. As a result, more recent verification techniques aim to detect anomalies in real-time. NetPlumber \cite{netplumber} was one of the earliest techniques in this direction. It creates a plumbing graph that identifies existing rules affected by the addition or deletion of new rules, and uses algebraic operations defined in HSA. However, for networks with a large number of rules, the plumbing graph can be large, resulting in high verification time.

Veriflow \cite{veriflow} addresses the challenge of dynamic network verification through the partitioning of packet headers into equivalence classes (EC) using a trie structure. It then checks network invariants by traversing a forwarding graph that corresponds to each EC. However, Veriflow's approach is limited to modeling forwarding functions in a network and only considers each EC in isolation, which can result in higher verification times.

Yang et al. \cite{atomic-predicate} presented an algorithm for verifying network properties in real-time using atomic predicates -- small and reusable building blocks that can express complex network properties. All predicates in AP Verifier are represented by binary decision diagrams (BDDs), which are rooted, directed acyclic graphs. Logical operations on BDDs can be performed efficiently using graph-based algorithms. 

Deltanet \cite{delta-net} builds on Veriflow's approach by constructing a single forwarding graph that covers all the equivalence classes (ECs) and then uses it to check network invariants. However, Deltanet's approach is limited to verifying reachability in the presence of forwarding rules and does not support modeling of other network functions, such as packet filtering and transformation. Deltanet also does not support batch processing. Vercel on the other hand, does batch processing, evaluates what-if conditions and tends to be more scalable due to vector algebra. 

APKeep \cite{apkeep} enhances the methods used in Veriflow and Deltanet by enabling the modeling of a wide range of network functions. This is accomplished by partitioning packet headers into equivalence classes (ECs) and representing network functions with Boolean formulas. The algorithm then verifies network invariants by solving these formulas using binary decision diagrams (BDDs). However, APKeep is only able to detect configuration errors, whereas Vercel can identify configuration errors and offer recommendations to correct them.
}

Katra \cite{beckett2022katra} uses pushdown systems for evaluating reachability in multi-layer networks. 
In contrast, Vercel models packet headers in a vector space and checks network invariants and policies by using least squares.
Flash \cite{flash} proposes an automata-theory-based solution to handle data plane verification amidst update storms and too-slow arrivals of update messages. 
Vercel performs well while handling update storms and outperforms state-of-the-art sequential data plane verification approaches.
Mahjong \cite{Mahjong} is a tool that helps users to choose among multiple dataplane verification approaches. Vercel is naturally suited to fast-handle storm of updates because of its scalability and vector-based parallel processing abilities.


\section{Summary}\label{sec:conclude}
We presented Vercel, inspired by the techniques of linear algebra to check reachability and delve into error rectification and recommendation. 
Vercel works by making use of vector spaces that are the result of mapping packet headers onto a binary tree.
Then, these vector spaces lead to the formation of a matrix $\mathcal{A}$, which represents the set of headers at a port, and vector $b$ -- the set of headers that need to be evaluated, resulting in $\mathcal{A}x=b$.
Since $\mathcal{A}x=b$ is not always solvable, Vercel deploys least squares, which guarantees a solution irrespective of the rank of $\mathcal{A}$.
Representing headers in vector space helps process multiple headers simultaneously. The use of vector algebra makes Vercel achieve aspects of verification like batch updates, what-if conditions, path and table recommendations beyond what other techniques can do.
Based on the experiments with real-world datasets, we show Vercel models a variety of network functions and checks for reachability and network invariants (loops, blackhole).

We show that Vercel is at least 70\% faster than the state-of-art techniques while providing more expressive power.
We also showed that Vercel could be used to evaluate reachability post link failures and check for routing policies.
The least squares solution also results in a recommendation model that checks tables for configuration anomalies to avoid longer paths. Vercel's vector space architecture leads to a paradigm shift -- where it takes in intents and converts these into table entries leading to automatic service provisioning.
The recommendation model provides for more possibilities beyond what we have so far studied in the domain of network verification. 

\bibliographystyle{IEEEtran}
\bibliography{references}

\appendix

\section{Proofs}\label{Appendix:proof}

\subsection{Proof of Theorem 3}\label{Theorem3proof}
\textit{Proof overview}: 
We start the proof by showing that if columns of the matrix $\mathcal{A}$ are selected from the standard basis (which is an orthonormal basis), then the projection point $b_i$ can be obtained in linear time by multiplying diagonal elements of the matrix $\mathcal{A}\mathcal{A}^T$ and vector $b_{i-1}$.
Thereafter, we show two cases with quadratic and linear time complexity to obtain the diagonal elements of the matrix $\mathcal{A}\mathcal{A}^T$.
This is a significant improvement compared to the use of only orthogonal vectors that require matrix product ($\mathcal{O}(m^{2.37})$ \cite{matmul}).

\textit{Proof}: 
With the standard basis, it is possible to reduce the time complexity of solving the linear equations $\mathcal{A}^p_ix^p_i=b_{i-1}$ and obtain the projection point $b_i$.
Now the columns of matrix $\mathcal{A}_i^p$ are orthonormal ($\mathcal{A}\mathcal{A}^T=I$), therefore the computation of a projection point $b_i$ from $b_{i-1}$ can be simplified as $b_i = \mathcal{A}^p_i(\mathcal{A}^p_i)^Tb_{i-1}$.
Note that the matrix product $\mathcal{A}^p_i(\mathcal{A}^p_i)^T \in \mathbb{R}^{m\times m}$ results in a diagonal matrix, whose $j\textsuperscript{th}$ diagonal entry will be ``1", if $j$\textsuperscript{th} row of the matrix is non-zero.
Otherwise, diagonal entries of $\mathcal{A}^p_i(\mathcal{A}^p_i)^T$ will be 0.
Therefore, vector $b_i$ is essentially an element-wise product between the diagonal elements of $\mathcal{A}^p_i(\mathcal{A}^p_i)^T$ and $b_{i-1}$.

Now, we ask if it is possible to create a vector $v_i^p$ (having $m$ elements) containing the diagonal elements of the matrix $\mathcal{A}^p_i(\mathcal{A}^p_i)^T$ but without performing the matrix product $\mathcal{A}^p_i(\mathcal{A}^p_i)^T$?
If we can create $v_i^p$ in linear time, then the projection point $b_i$ can be computed efficiently (linear time) as $b_i = v_i^p \otimes b_{i-1}$, where $\otimes$ is the element-wise product between two vectors.
There exist two possibilities for creating $v_i^p$ (without computing $\mathcal{A}^p_i(\mathcal{A}^p_i)^T$), each leading to different complexities.

\textit{Case 1: Modeling forwarding behavior with the complexity of $\mathcal{O}(m^2)$}.
Represent $v_i^p$ as the sum of columns of $\mathcal{A}_i^p$ i.e. $(v_i^p)^j = \sum_{k=1}^{m}(\mathcal{A}_i^p)^{jk}$.
In this case, complexity of creating $v_i^p$ is $\mathcal{O}(m^2)$ and therefore, complexity for obtaining the projection point $b_i$ from $b_{i-1}$ is $\mathcal{O}(m^2)$.

\textit{Case 2: Modeling forwarding behavior with the complexity of $\mathcal{O}(m)$}.
Instead of obtaining vector $v_i^p$ from matrix $\mathcal{A}_i^p$, create $v_i^p=\vec{0}$ in $m$-dimensional space.
Subsequently, if router $i$ forwards $j\textsuperscript{th}$ header (in the set $S\textsuperscript{affected}$) to its output port $p$, then update the $j\textsuperscript{th}$ entry of $v_i^p$ to 1.
After updation, the forwarding vector $v_i^p$ denotes atomic+iatomic headers (in the set $S\textsuperscript{affected}$) that router $i$ forwards to its output port $p$.
Note $v_i^p$ still represents the sum of columns of the forwarding matrix $\mathcal{A}_i^p$.
However, the complexity of updating $v_i^p$ is $\mathcal{O}(m)$ because in the worst case, each entry of the vector is visited only once.
Hence, complexity for obtaining the projection point $b_i$ from $b_{i-1}$ is $\mathcal{O}(m)$.
This proves the theorem.

\subsection{Proof of Theorem 4}\label{Theorem4proof}
The representation of headers in the set $S\textsuperscript{affected}$ using a standard basis also reduces the complexity of finding the reachable headers at the destination node.
In the existing solution (Section \ref{sec:design}), an atomic+iatomic header $s_k\in S$ is reachable between a source and destination if the dot product between $b_{reachable}$ and the vector notation ($e_k$) of header $s_k$ is non-zero.
By performing dot product, we can obtain all reachable atomic+iatomic headers with the complexity of $\mathcal{O}(m^2)$ (each dot product is of complexity $\mathcal{O}(m)$) and we need to perform $m$ such dot products.
However, suppose we utilize the standard basis vectors to represent atomic+iatomic headers. 
In that case, all reachable headers correspond to the indices with non-zero entries of $b_{reachable}$ and the complexity of finding non-zero entries in an $m$-dimensional vector is $\mathcal{O}(m)$.
This completes the proof of the theorem.

\subsection{Proof of Theorem 5}\label{Theorem5proof}
After a rule update (insertion/deletion), Vercel performs the following three steps to compute reachability along a path: 1. Find the sets $S\textsuperscript{affected}$ and $P\textsuperscript{affected}$; 2. Model forwarding behavior of routers' ports present in the path; 3. Determine reachable headers at the destination from the vector space.
From Theorem 2, we know that the time complexity of finding the sets $S\textsuperscript{affected}$ and $P\textsuperscript{affected}$ is $\mathcal{O}(m)$. Thereafter, Theorem 3 helps to model the forwarding behavior of a router with time complexity of $\mathcal{O}(m)$. 
Now, consider that the number of routers along a path are constant ($c$) w.r.t. to the number of atomic+iatomic headers in a network. 
In this case, the complexity to compute reachability along the path is $\mathcal{O}(c*m)=\mathcal{O}(m)$. 
Finally, Theorem 4 provides a solution (of complexity $\mathcal{O}(m)$) to recover reachable headers from the vector space. 
Note that each step (steps 1-3 described above) requires $\mathcal{O}(m)$ time. 
Therefore, Vercel solves reachability between a source and destination along a path with time complexity of $\mathcal{O}(m)$. 
This completes the proof of the theorem.

\subsection{Proof of Theorem 6}\label{Theorem6proof}
As described in the proof presented in Appendix \ref{Theorem4proof}, if we represent atomic+iatomic headers using standard basis vectors, then it is possible to create a forwarding vector $v_i^p\in \mathbb{R}^m, m=|S\textsuperscript{affected}|$ representing the atomic+iatomic headers that router $i$ forwards along its port $p$.
Since, a forwarding vector $v$ is created for each port in the set $P\textsuperscript{affected}$, therefore Vercel requires $|S\textsuperscript{affected}|*|P\textsuperscript{affected}|$ bits to create the forwarding vectors $v_i^p$.
In other words, the space complexity of Vercel (by using the forwarding vectors $v_i^p$) is $\mathcal{O}(|S\textsuperscript{affected}|*|P\textsuperscript{affected}|)$.
For example, consider a network with 10K atomic+iatomic headers and 200 ports across the network. 
If after a rule insertion, $|S\textsuperscript{affected}|=1$K and $|P\textsuperscript{affected}|=50$, then Vercel requires $\sim6$ KB to represent the forwarding vectors $(v_i^p)$.
This completes the proof of the theorem.

\end{document}